\documentclass[twocolumn,preprintnumbers,amsmath,amssymb]{revtex4}  

\usepackage{amsfonts}
\usepackage[T1]{fontenc}
\usepackage{amsmath,amsbsy,amssymb,graphicx}
\usepackage{times}
\usepackage{amsmath}
\usepackage{amssymb}
\usepackage{graphicx}
\usepackage{color}

\begin{document}

\title{Layer-by-layer assembly of multilayer optical lattices: Application to displaced dice lattice}

\author{Lei Hao}
 \address{School of Physics, Southeast University, Nanjing 211189, China}

\date{\today}

\begin{abstract}
We propose methods for synthesizing multilayer optical lattices of cold atoms in a layer-by-layer manner, to unlock the potential of optical lattices in simulating the fascinating physics of multilayer systems. Central to the approach is to compress the beam profile of a red-detuned Gaussian laser beam from disklike to a thin line by a telescope with two cylindrical lenses. A highly tunable multilayer optical lattice is obtained by passing the compressed Gaussian beam through an optical device consisting of beam splitters, mirrors, and glass plates. We illustrate the proposal with the displaced dice lattice, which is a trilayer lattice that maps to the dice lattice when projected to the same layer. Both the dice model and its interesting variants may be realized. For a model of fermionic cold atoms, featuring an isolated flat band between two dispersive bands, we find valley-contrasting interband transitions involving the flat band.
\end{abstract}


\maketitle

\section{Introduction}

Optical lattices of cold atoms have become a fertile ground to simulate novel quantum systems and to explore fundamental open problems in physics \cite{lewenstein07,bloch08,dalibard11,zhang18,cooper19}. In the past 30 years, a great variety of optical lattices, ranging from zero dimensional quantum dots to three dimensional (3D) lattices, have been studied both theoretically and experimentally \cite{petsas94,grynberg01,grimm00,windpassinger13}.
Motivated partly by the keen interest from the condensed matter community, the two-dimensional (2D) optical lattices have attracted special attention. Besides purely 2D materials containing a single atomic layer, such as graphene and its analogue \cite{novoselov04,zhang10nphys}, the majority of 2D systems studied in condensed matter physics consist of several atomic layers and are actually multilayer (or, few-layer) quasi-2D (q-2D) materials. They host properties determined not only by the constituent layers but also by their stacking arrangement, thereby breeding a field rich in novel phases and striking phenomena. A notable example is the bilayer graphene, which opens a tunable band gap under a vertical electric field \cite{mccann06,castro07,min07}, and turns into a Mott insulator or unconventional superconductor after twisting the two layers by a magic angle \cite{cao18a,cao18b}. With the superior controllability of the optical lattices, it seems promising to fabricate multilayer optical lattices with cold atoms to explore physics far beyond the reach of present-day solid state physics. In reality, however, multilayer optical lattices are seldom studied \cite{uehlinger13,wu13,tao14,noda14,greif15,dey16,guardian16,grab16,tudela19,shen20,salamon20,fu20}.

Most existing studies of multilayer optical lattices rely on restricting the number of layers of a 3D optical lattice in one direction (e.g., along the $z$ axis) \cite{uehlinger13,wu13,tao14,noda14,greif15,dey16}. The geometry of individual layers and the stacking of consecutive layers are both fixed by the 3D optical dipole potential and not free to tune. This approach, extracting a q-2D multilayer lattice from a 3D lattice, is analogous to the mechanical exfoliation in condensed matter physics \cite{novoselov04,li13}. Another powerful technique in condensed matter physics is the molecular-beam epitaxy (MBE), in which a multilayer lattice is synthesized layer by layer. The MBE does not require the existence of a bulk material consisting of weakly coupled 2D or q-2D units and is suitable for producing multilayers of both single crystals and complicated heterostructures \cite{he13,bollinger16,suyolcu20}.

The above comparison inspires us to explore an MBE-type layer-by-layer scheme of fabricating multilayer optical lattices. We illustrate in what follows that the layer-by-layer scheme for a multilayer optical lattice can be implemented through a combination of three sets of optical elements: The \emph{beam shapers} to compress the beam profile (i.e., cross section) of the input Gaussian beam from disklike to a thin line, the \emph{beam splitters} to split a single laser beam into several beams, and the \emph{path locators} to guide different laser beams to the spot of the optical lattice along appropriate paths. The beam splitters and path locators also produce relative phase shifts and play the role of \emph{phase shifters}. Fig. 1(a) is a schematic plot of the proposal.

The remaining part of the paper is organized as follows. In Sec.II, we explain the essential factors to materialize this proposal. Then, in Sec.III, we expound the proposal in more details in terms of a trilayer optical lattice, the displaced dice (or, $\mathcal{T}_{3}$) lattice. In Sec. IV we continue to show that the dice model and its close variants may be simulated by fermionic cold atoms in the displaced dice lattice. In a symmetrically biased dice model, which has an isolated flat band in between two dispersive bands, we find valley-contrasting interband transitions between the flat band and one dispersive band. The main results of the study are summarized in Sec. V. We also discuss in this section possible generalizations of the scheme to other novel multilayer lattices, such as the twisted multilayers. In order to focus on elucidating the physical pictures in the main text, we have put some relevant mathematical details to the Appendices (A through F).

\section{general idea of the layer-by-layer scheme}

\begin{figure}[!htb]\label{fig1}
\centering
\hspace{-5.5cm} {\textbf{(a)}}\\
\includegraphics[width=7.5cm,height=1.62cm]{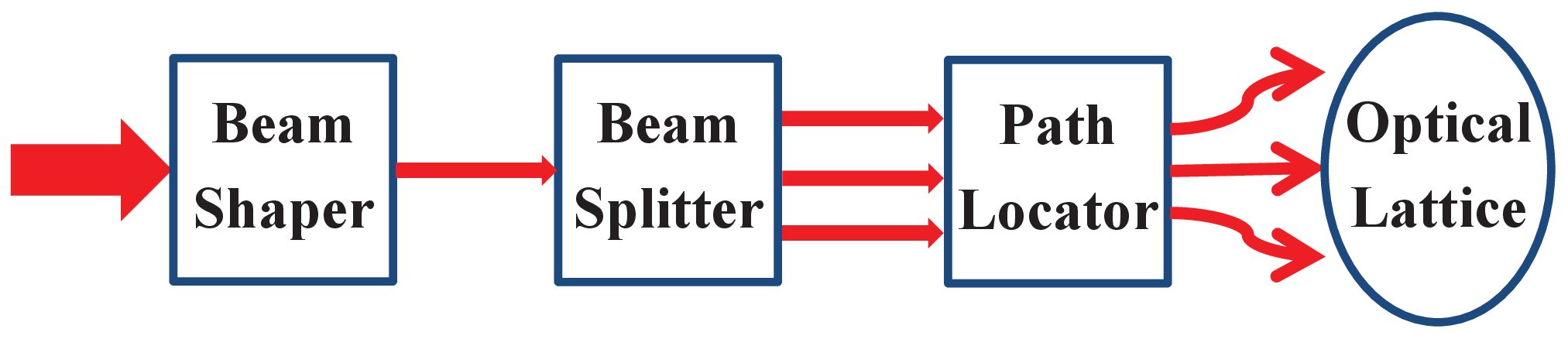} \\ \vspace{-0.05cm}
\hspace{-5.5cm} {\textbf{(b)}}\\
\includegraphics[width=6.5cm,height=2.913cm]{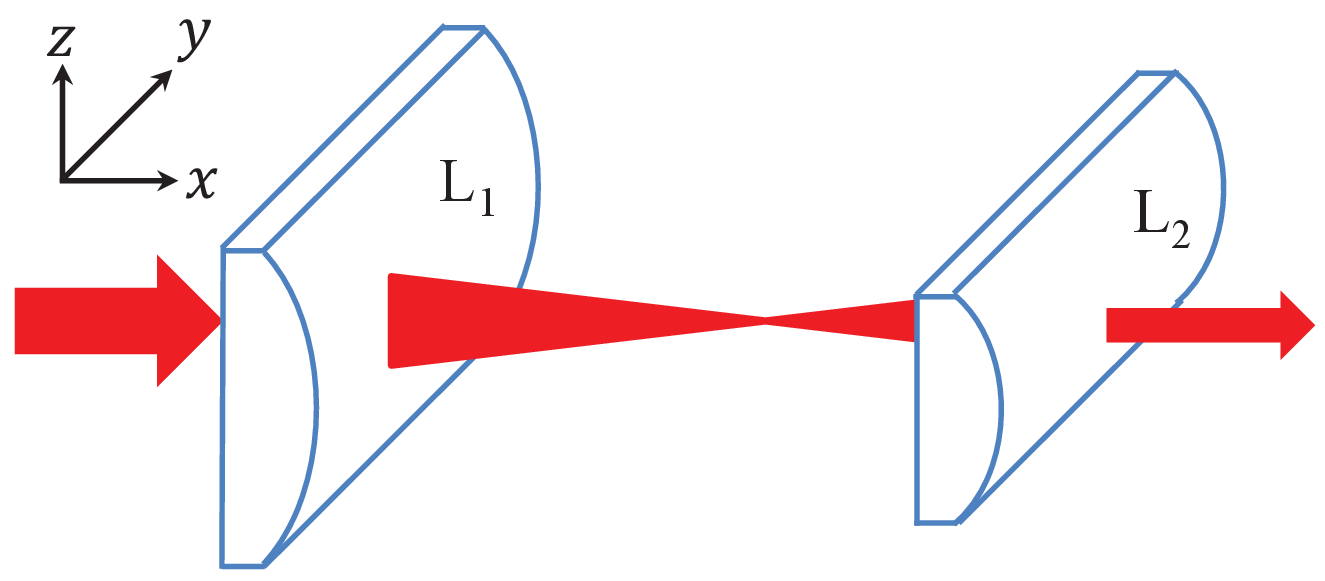}  \\ \vspace{-0.05cm}
\hspace{-5.5cm} {\textbf{(c)}}\\
\includegraphics[width=6.5cm,height=1.605cm]{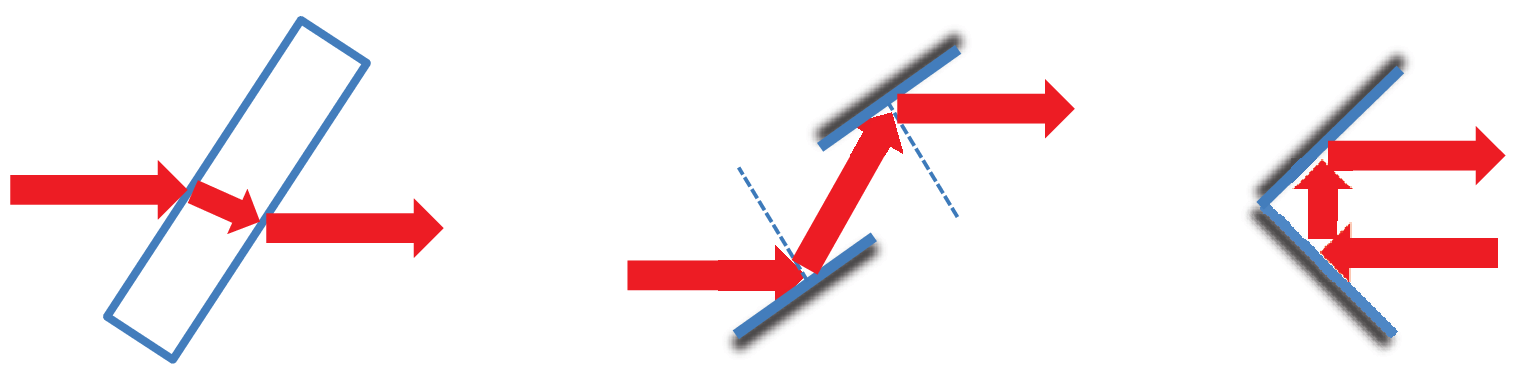}  \\
\caption{(a) Schematic of an optical device transforming the incoming Gaussian laser beam into the multilayer optical lattice, in a layer-by-layer manner. (b) The beam shaper in terms of a cylindrical Keplerian telescope consisting of two cylindrical thin lenses. It compresses the incoming laser beam along the $z$ direction. (c) Path locators that cause a lateral shift to the laser beam. The left is a glass plate whose normal is tilted from the propagation direction of the laser beam. The middle consists of two narrow mirrors parallel to each other. The right is a special Fresnel bimirror whose intersection angle is $\frac{\pi}{2}$. The propagation direction of the laser beam is unchanged (reversed) in the left and middle (right).}
\end{figure}

The key prerequisite of realizing the layer-by-layer assembling of multilayer optical lattices, in analogy to the MBE for condensed matter systems, is to generate individual purely 2D optical lattices.
Usually, a 2D optical lattice is formed either by enhancing the depth of a 3D optical potential (or, by adding an additional trapping potential) along the third direction to pinch off the interlayer couplings and focus on a single layer of the 3D lattice \cite{greif16}, or by strongly constraining the atoms only in the in-plane directions and obtain a 2D lattice of tube-like quasi-1D units which extend much longer than the in-plane lattice constant \cite{ozawa17}. These approaches to a 2D optical lattice are clearly inapplicable to the layer-by-layer scheme for the multilayer optical lattices.

Therefore, the beam shaper in Fig. 1(b), which compresses the input Gaussian laser beam to the output beam with a straight filamentary beam profile, plays a central role in realizing \emph{truly 2D} optical lattices of point-like lattice sites and in implementing the layer-by-layer assembling of multilayer optical lattices. The beam shaper in Fig.1(b) is a cylindrical Keplerian telescope consisting of a pair of cylindrical thin lenses. A cylindrical Galilean telescope applies equally well. We assume an incident laser beam propagating along the telescope's optical axis, which is parallel to the $x$ axis of the coordinates defined in Fig. 1(b). This beam shaper does not change the $y$-axis distribution of the input laser beam, but compresses its $z$-axis distribution \cite{sveltobook,siegmanbook}. Defining the focal lengths of the left (L$_{1}$) and right (L$_{2}$) cylindrical lenses as $f_{1}$ and $f_{2}$ ($f_{1}>f_{2}$), the waist size for the $z$-axis amplitude distribution is reduced from the original value $\rho_{0}$ to $\rho_{0z}=\rho_{0}f_{2}/f_{1}$ (see Appendix A for relevant formulas). $f_{1}/f_{2}$ is the telescope's magnification along $z$ direction. The larger the magnification is, the thinner the slab-like outgoing laser beam.

The waist radius $\rho_{0}$ of the input Gaussian beam is in the order of 1 mm \cite{sveltobook}.
To get the 2D limit, while it appears necessary to reduce the thickness of the slab-like output beam to approximately 1 $\mu$m, comparable to the in-plane lattice parameters, a thickness of about 10 $\mu$m is actually small enough for sufficiently deep optical potentials (see Appendix B). This amounts to reducing $\rho_{0z}$ of the output laser beam to about 5 $\mu$m. When line beams of this thickness are used to make a realistically deep 2D optical lattice, the energy level spacing between the ground state and the first excited state of a single potential well can be made larger than the recoil energy $E_{R}=h^{2}/(2m\lambda^{2})$ (see Appendix B). In the low-temperature limit \cite{aspect88,kasevich92} and without strong driving, the resulting optical lattice can safely be considered as a truly 2D lattice. Compressing an input Gaussian beam with $\rho_{0}\approx1$ mm to $\rho_{0z}\approx5$ $\mu$m requires $f_{1}/f_{2}\approx200$. Besides using a single cylindrical Keplerian telescope with $f_{1}/f_{2}\approx200$, we may also cascade two cylindrical Keplerian (or, Galilean) telescopes with magnifications around 10 to 20. In the cascaded configuration, the whole device is much more compact.

After accomplishing the above step, the multilayer optical lattices may in principle be realized by passing the line beam through a proper combination of standard optical elements, such as beam splitters, glass plates, and mirrors. In Fig. 1(a), these are listed separately as Beam Splitter and Path Locator. They will usually be intertwined with each other in actual devices. According to the geometries of various layers and the number of layers, we split the line beam into several beams with equal or unequal intensities, by using beam splitters with suitable splitting ratios. For example, a triangular optical lattice may be formed by three traveling wave laser beams \cite{grynberg01}. To synthesize a q-2D optical lattice containing $N$ triangular layers, we split the single line beam into $3N$ beams. Guiding the various laser beams to a designated interference region, the multilayer optical lattice is created therein. The distances between consecutive layers may be controlled by the path locators listed in Fig. 1(c).

To ensure the stability and accuracy of the whole optical device, it is advantageous to integrate the optical components therein within a single platform that allows precise control over both the positions and the orientations of all components. In addition, it is preferable to have all the optical surfaces in the device superpolished. Although technically demanding, there is no fundamental restriction preventing the realization of the proposal.

Loading precooled cold atoms to the multilayer optical lattice may follow either of two procedures. Firstly, we may produce the different layers of the optical lattice in sequence, and also load precooled cold atoms to these layers in a sequential manner, fully analogous to the MBE. Secondly, we may produce all the layers of the optical lattice at the same time but distant enough from each other. Then we load cold atoms to the detached layers and move them towards each other along the vertical direction. In both approaches, the distances between neighboring layers and their relative orientations should be controlled precisely in concord. The atoms loaded to different layers could be the same or different, fermionic or bosonic. When the same species of cold atoms are loaded to all layers, we may directly bring all the layers in place and then load the cold atoms to them simultaneously.

It is now imperative to demonstrate through a concrete example the feasibility and fine details of the proposal and the promised tunability of the model parameters for the target system. For this purpose, we consider the displaced dice (or, $\mathcal{T}_{3}$) lattice. The dice lattice and its variants (e.g., the $\alpha$-$\mathcal{T}_{3}$ lattice) have been extensively studied theoretically in both the cold atoms community and condensed matter physics \cite{sutherland86,vidal98,rizzi06,bercioux09,wang11,xu17,ocampo17,illes15}. While the dice lattice has been fabricated in superconducting networks \cite{abilio99,serret02} and normal metal networks \cite{naud01}, the dice model may not be realized in them. Hopefully, the present proposal will facilitate the experimental study of the pure and the generalized dice models. The following analysis for the displaced dice lattice may easily be adjusted to apply to other multilayer optical lattices.

\section{Displaced dice lattice: The optical lattice}

As shown in Fig.2, the (displaced) dice lattice consists of three sublattices (A, B, and C) which separately form a triangular lattice \cite{sutherland86,vidal98,rizzi06,bercioux09}. In the limit that the dice model or its close variants (e.g., the $\alpha$-$\mathcal{T}_{3}$ model) are applicable, only the bonds connecting the atoms on the A and C sublattices (the rim sublattices) and the nearest-neighboring (NN) atoms of the B sublattice (the hub sublattice) are retained. Here, we propose to stagger the three sublattices along the direction perpendicular to the layer planes, so that the A and C layers are separately above and below the B layer, at the same or different distances from the B layer.

\begin{figure}\label{fig2} \centering
\includegraphics[width=6.5cm,height=6.666cm]{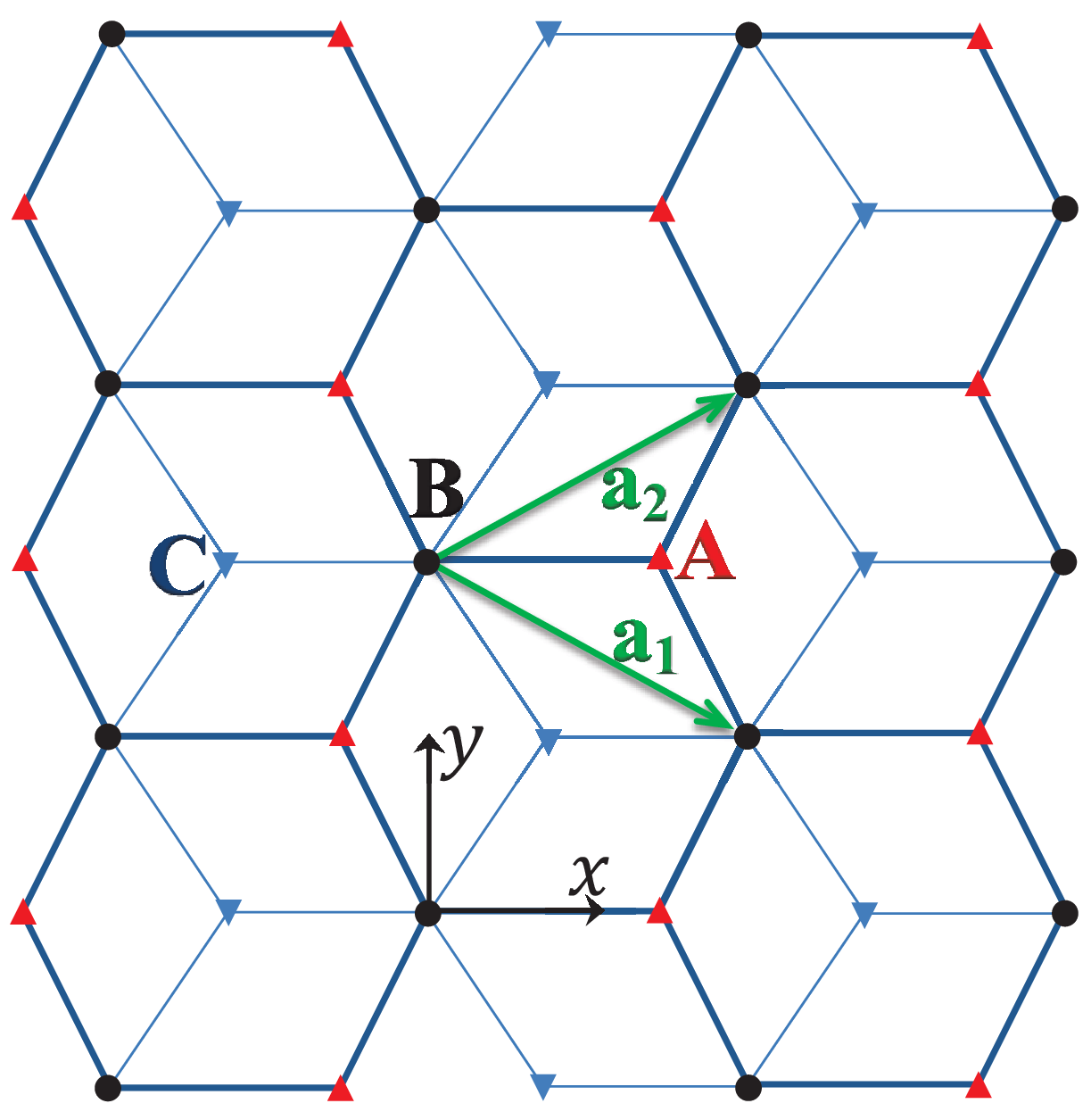}
\caption{The dice ($\mathcal{T}_{3}$) lattice. The A, B, and C sublattices are marked by red upward triangles, black circles, and blue downward triangles. The two arrowed green lines $\mathbf{a}_{1}=(\frac{\sqrt{3}}{2},-\frac{1}{2})a$ and $\mathbf{a}_{2}=(\frac{\sqrt{3}}{2},\frac{1}{2})a$ are the primitive lattice vectors. In the displaced lattice, the plane of the A (C) sublattice is above (below) the plane of the B sublattice.}
\end{figure}

Each triangular layer may be formed by superimposing three traveling wave laser beams \cite{petsas94,grynberg01,becker10}. Therefore, we need nine laser beams to construct the displaced dice lattice. We denote the wave vectors of the three laser beams for the $i$-th layer as $\mathbf{k}^{(j)}_{i}$ ($i,j=1,2,3$). Since the three triangular layers of the displaced dice lattice have the same primitive lattice vectors $\mathbf{a}_{1}$ and $\mathbf{a}_{2}$, we choose $\mathbf{k}^{(1)}_{i}=k(\sqrt{3}/2,-1/2,0)=\mathbf{k}_{1}$, $\mathbf{k}^{(2)}_{i}=k(0,1,0)=\mathbf{k}_{2}$, and $\mathbf{k}^{(3)}_{i}=k(-\sqrt{3}/2,-1/2,0)=\mathbf{k}_{3}$ for $i=1,2,3$. For a laser of wavelength $\lambda$, the wave number $k=2\pi/\lambda$. The electric fields of the nine laser beams are written as
\begin{equation}
\mathbf{E}_{i}^{(j)}(\mathbf{r},t)=E_{i}\hat{\boldsymbol{\epsilon}}_{j}\cos(\mathbf{k}_{j}\cdot\mathbf{r}-\omega t+\phi_{i}^{(j)})g_{i}^{(j)}(\mathbf{r}).
\end{equation}
$g_{i}^{(j)}(\mathbf{r})$ describes the beam profile of the compressed Gaussian beams \cite{sveltobook}. We assume that the cold atoms are loaded to the middle of the whole optical lattice, in a region characterized by $\rho_{0z}$ (i.e., $2\pi\rho_{0z}^{2}/\lambda\simeq10\pi\rho_{0z}$ for $\rho_{0z}\simeq5\lambda$) and much narrower than the width of the line beam (i.e., $\sim2\rho_{0}$) (see Appendix A). In this case, the variation of $g_{i}^{(j)}(\mathbf{r})$ along the width of the line beam is negligible. It is then a good approximation to take
\begin{equation}
g_{i}^{(j)}(\mathbf{r})\rightarrow g_{i}(z)\approx e^{-(z-z_{i})^{2}/\rho_{0z}^{2}},
\end{equation}
suppose the region is also centering around the waist of the $z$-axis amplitude distribution. $z_{i}$ is the center of the $i$-th layer along the $z$ axis. To be specific, we label the A, B, and C layers in sequence by $i=$1, 2, and 3.

Setting the polarizations $\hat{\boldsymbol{\epsilon}}_{j}=\hat{\mathbf{z}}$ ($j=1,2,3$), the optical dipole potential for the $i$th layer is \cite{petsas94,grynberg01,becker10}
\begin{equation}
U_{i}(\mathbf{r})=-\frac{\epsilon_{0}}{4}\alpha_{0}^{\prime}E_{i}^{2}(\mathbf{r}),
\end{equation}
where
\begin{eqnarray}
&&E_{i}^{2}(\mathbf{r})=E_{i}^{2}g_{i}^{2}(z)\{3+2[\cos(\mathbf{k}_{12}\cdot\mathbf{r}+\phi_{i}^{(12)})  \notag \\
&&\hspace{0.5cm}  +\cos(\mathbf{k}_{23}\cdot\mathbf{r}+\phi_{i}^{(23)})+\cos(\mathbf{k}_{13}\cdot\mathbf{r}+\phi_{i}^{(13)})]\}.
\end{eqnarray}
$\mathbf{k}_{ij}=\mathbf{k}_{i}-\mathbf{k}_{j}$, $\phi_{i}^{(mn)}=\phi_{i}^{(m)}-\phi_{i}^{(n)}$. $\alpha_{0}^{\prime}$ is the real part of the polarizability. For red-detuned laser, $\alpha_{0}^{\prime}>0$, the minima of the optical dipole potential reside at the maxima of $E_{i}^{2}(\mathbf{r})$ \cite{grynberg01}. For $^{40}$K \cite{mckay11}, red-detuned lasers with wavelength at 1064 nm \cite{jordens08,tarruell12,greif13} and 1030 nm \cite{schneider08,hackermuller10} were used in experiments.

When the three layers are isolated (i.e., $z_{1}-z_{2}\gg\rho_{0z}$ and $z_{2}-z_{3}\gg\rho_{0z}$), the full optical potential $V(\mathbf{r})$ is the sum of Eq.(3) over the three layers. As we move the three layers closer so that laser beams for NN layers start to overlap, interference between the laser beams of NN layers adds new interlayer interference terms to $V(\mathbf{r})$. While these interlayer interference terms are interesting, we will neglect them in the following analysis and defer more serious consideration of their effects to later studies. In experiments, these interlayer interference terms may be eliminated by making the laser frequencies for different layers slightly different \cite{tarruell12,tarruell18}, through nonlinear optical effects such as the stimulated Raman scattering or optical parametric oscillation \cite{hitzbook}. In this way, while the interlayer interference terms average out with time, the slight difference in frequency does not lead to appreciable differences in the lattice parameters of different layers. The full optical potential is therefore adequately described as the sum of the three single-layer components defined by Eqs.(3) and (4) \cite{tarruell12,tarruell18}.

By finely adjusting the phase parameters $\phi_{i}^{(j)}$, we may align the projections of the three triangular layers to the $xy$ plane according to Fig.2. One way is to set
$\phi_{i}^{(1)}=\phi_{1}+2(i-1)\theta$, $\phi_{i}^{(2)}=\phi_{2}+(i-1)\theta$, and $\phi_{i}^{(3)}=\phi_{3}$. $\phi_{i}$ ($i=1,2,3$) are arbitrary constants. $\theta$ determines the successive shift of the three triangular layers along the $x$ direction. When $\theta=2\pi/3$, we get the displaced dice lattice shown in Fig. 2. Of course, each phase may be different from the above value by an integral multiple of $2\pi$. The two reciprocal lattice vectors are $\mathbf{b}_{1}=\mathbf{k}_{12}$ and $\mathbf{b}_{2}=\mathbf{k}_{23}$. The lattice constant is related to the wavelength by $k=2\pi/\lambda=4\pi/(3a)$, which gives $a=2\lambda/3$.

\begin{figure}[!htb]\label{fig3}
\centering
\hspace{-5.5cm} {\textbf{(a)}}\\
\includegraphics[width=6.5cm,height=4.2cm]{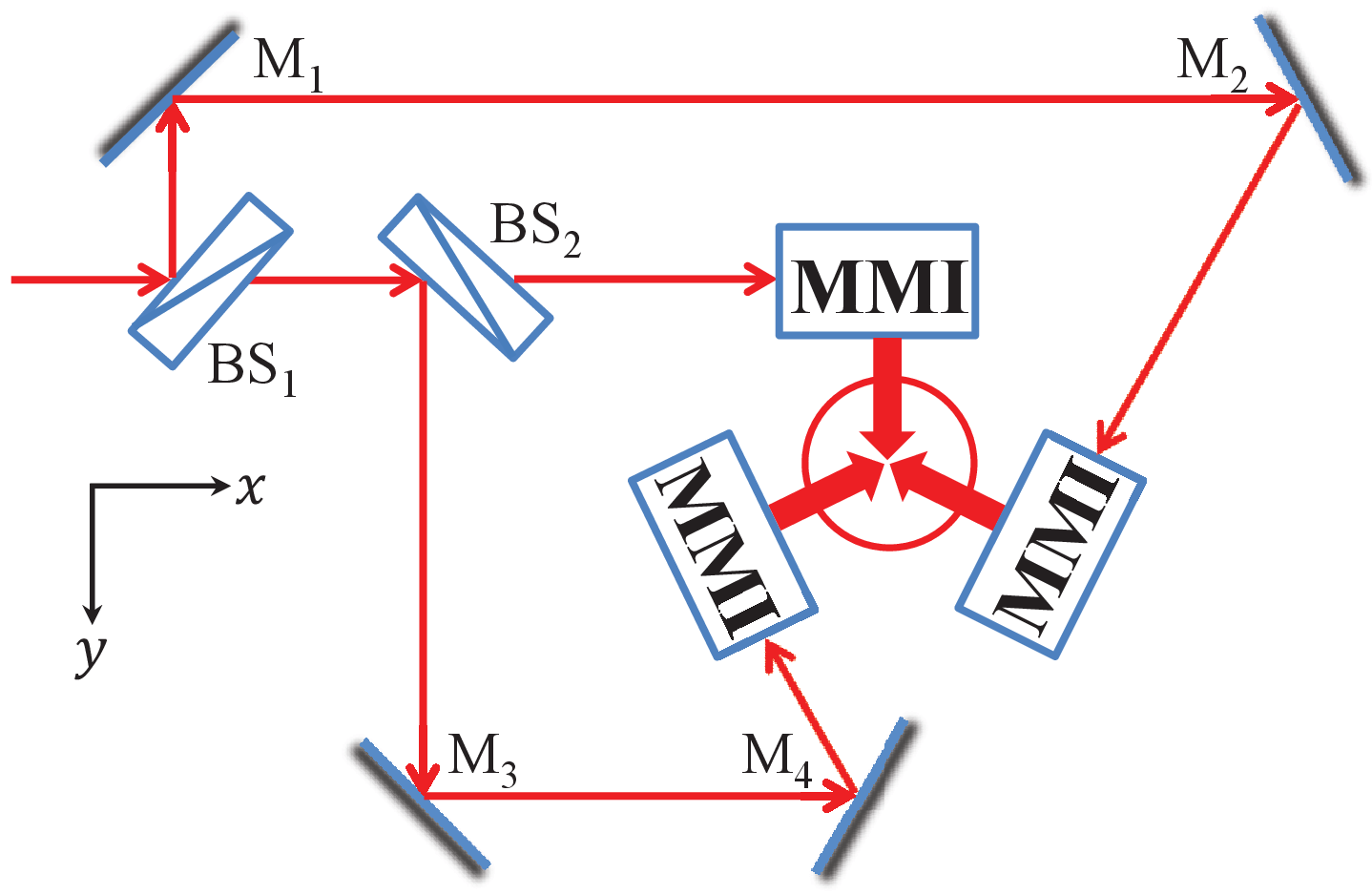} \\ \vspace{-0.05cm}
\hspace{-5.5cm} {\textbf{(b)}}\\
\includegraphics[width=6.5cm,height=3.458cm]{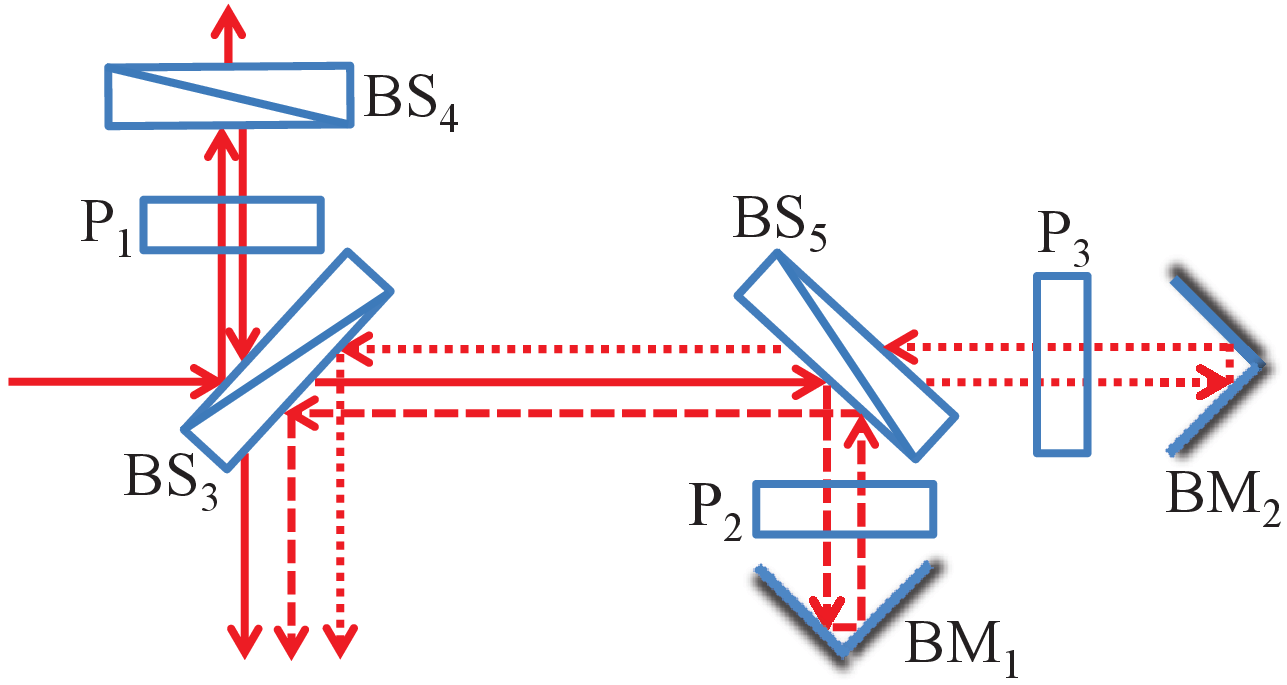}  \\
\caption{(a) One possible combination of the beam splitters and path locators, for the displaced dice lattice. MMI is the shorthand for the modified Michelson interferometer. The circle in the middle of three MMIs surrounds the optical lattice. (b) The structure of the MMI, which splits one input laser beam into three output beams and shifts them along the $z$ direction (i.e., perpendicular to the lattice planes). For clarity of illustration, the two bimirrors BM$_{1}$ and BM$_{2}$ have been rotated by $\pi/2$ from the actual configurations, with respect to the propagation direction of the corresponding incident laser beams. Correspondingly, the three outgoing beams are actually shifted along the $z$ direction rather than in the $xy$ plane. In the MMI, we assume the coating layer of BS$_{3}$ (BS$_{5}$) is on the right (left) surface. The splitting ratios may vary from BS$_{1}$ to BS$_{5}$, to control the relative amplitudes of different beams.}
\end{figure}

To realize the displaced dice lattice defined above, we pass the line beam output from the cylindrical Keplerian telescope through the optical device in Fig. 3 (see also Appendix D for another optical device). After splitting the line beam into three beams by the beam splitters BS$_{1}$ and BS$_{2}$ in Fig. 3(a), we inject each of the three beams into a compound optical element that we call the modified Michelson interferometer (MMI). The MMI shown in Fig.3(b) splits the incident laser beam into three parallel beams that are separated along the $z$ direction but coincide with each other when projected along the $z$ direction onto the $xy$ plane. The vertical distances between consecutive layers are controlled by the special Fresnel bimirrors with 90$^{\circ}$ intersection angle between the two mirrors, see also Fig.1(c). These bimirrors may also be replaced by right-angle prisms made of glass with a high refractive index (see Appendix C). Three glass plates (P$_{1}$ to P$_{3}$) are inserted to the light paths, to act as phase shifters. Besides using plates of varying thickness, we may also tilt the plates to control the phase shifts. If frequency modulation is required, we insert a suitable nonlinear optical element to the corresponding light path \cite{hitzbook}.

\section{Tight-binding models for fermionic cold atoms in the displaced dice lattice}

Instead of exploring the full parameter ranges of the displaced dice lattice, we fix $\mathbf{k}_{i}^{(j)}$ and $\phi_{i}^{(j)}$ ($i,j=1,2,3$) to the ideal values defined in the previous section. We also consider loading the same fermionic cold atoms, such as $^{40}$K or $^{6}$Li, to all three layers. The polarizations of the laser beams will be fixed to $\hat{\boldsymbol{\epsilon}}_{j}=\hat{\mathbf{z}}$ ($j=1,2,3$). In addition, the laser beam output from the cylindrical telescope has a fixed waist size $\rho_{0z}$ along the $z$ axis. Under these restrictions, the remaining tunable parameters of the optical lattice include $E_{i}$ and $z_{i}$ ($i=1,2,3$). Taking $E_{2}$ and $z_{2}=0$ for the B layer as references, and the remaining four parameters as free parameters, we are able to simulate a rich family of interesting models.

Assuming all three layers have attained the 2D limit, and the ultracold atoms only occupy the lowest energy levels of the various potential wells. This justifies a tight-binding description, since these orbitals are deeply localized in the potential wells. For our displaced dice lattice, the single-orbital tight-binding model up to NN hopping has nine parameters, including three on-site energies, three intralayer NN hopping amplitudes, and three interlayer NN hopping amplitudes. A qualitative estimation of the parameters follows by making the harmonic approximation to the potential wells and taking the lowest eigenstates as the local Wannier orbitals (see Appendix E). The dice model and its variants require the AB and BC interlayer hopping amplitudes to dominate over the three intralayer and the AC interlayer hopping amplitudes \cite{sutherland86,vidal98}. This is achieved in the present layer-by-layer scheme through the exponential decay of the NN hopping amplitudes with the relevant inter-site distances \cite{lewenstein07,bloch08}. Namely, according to Fig.2, the AB and BC interlayer hopping amplitudes may dominate over the intralayer hopping amplitudes because the $xy$ projection of the NN distance is $a/\sqrt{3}$ for the former versus $a$ for the latter. The AB and BC NN interlayer hopping amplitudes may dominate over the AC NN interlayer hopping amplitude because the projection of the NN bonds to the $z$ direction for AC, $z_{1}-z_{3}$, is about twice as large as $z_{1}-z_{2}$ and $z_{2}-z_{3}$ for AB and BC (i.e., in most cases of interest we have $z_{1}-z_{2}\simeq z_{2}-z_{3}$). Therefore, by finely tuning the $z$ coordinates of the three layers, we may make the AB and BC NN interlayer hopping amplitudes dominate over the others (see Appendix E). Together with the control over $E_{1}/E_{2}$ and $E_{3}/E_{2}$, we may arrive at the desired dice model and its variants.

In conclusion, by tuning four free parameters ($E_{1}/E_{2}$, $E_{3}/E_{2}$, $z_{1}$, $z_{3}$), we may realize the following family of four-parameter tight-binding models
\begin{eqnarray}
\hat{H}&=&\sum\limits_{\langle i,j\rangle,\sigma}(t_{ba}b^{\dagger}_{i\sigma}a_{j\sigma}
+t_{bc}b^{\dagger}_{i\sigma}c_{j\sigma}+\text{H.c.})   \notag \\
&&+\sum\limits_{i,\sigma}(\varepsilon_{ab}a^{\dagger}_{i\sigma}a_{i\sigma}
+\varepsilon_{cb}c^{\dagger}_{i\sigma}c_{i\sigma}).
\end{eqnarray}
The summation $\langle i,j\rangle$ runs over NN intersublattice sites. The index $\sigma$ denotes the two spin states of the fermionic cold atoms. H.c. means the Hermitian conjugate of the terms explicitly written out. We have taken the on-site energy for the B layer as reference, so $\varepsilon_{ab}=\varepsilon_{a}-\varepsilon_{b}$ and $\varepsilon_{cb}=\varepsilon_{c}-\varepsilon_{b}$.
For $\varepsilon_{ab}=\varepsilon_{cb}=0$ and $t_{bc}=t_{ba}$, we get the dice ($\mathcal{T}_{3}$) model \cite{sutherland86}. For $\varepsilon_{ab}=\varepsilon_{cb}=0$ and $t_{bc}=\alpha t_{ba}$ ($0<\alpha<1$) we get the $\alpha$-$\mathcal{T}_{3}$ model \cite{illes15}. These two families of models both have a flat band connecting linearly to one upper and one lower dispersive bands. For $\varepsilon_{ab}=\varepsilon_{cb}\neq0$ and $t_{bc}=\alpha t_{ba}$ ($0<\alpha\le1$), the flat band connects quadratically only to one dispersive band \cite{wang11}. We define the displaced dice lattice as \emph{biased} if $\varepsilon_{ab}\varepsilon_{cb}<0$, in analogy to a semiconductor slab under an electric field perpendicular to the slab, i.e. an electric bias \cite{mccann06,castro07,min07}. The inversion symmetry is broken in the biased dice lattice.
In previous proposals for the dice lattice \cite{rizzi06,bercioux09}, the symmetry of the optical potential allows only the parameters for the pure dice model or $\varepsilon_{ab}=\varepsilon_{cb}\neq0$ and $t_{bc}=t_{ba}$. This comparison highlights the flexibility of the multilayer optical lattices obtained by the layer-by-layer approach.

%
%

To further motivate the above highly tunable model, we study the interband transitions in a biased dice model for $t_{ba}=t_{bc}=t_{0}$ and $\varepsilon_{ab}=-\varepsilon_{cb}=\Delta$. Fermionic cold atoms in this model features a band structure with a flat band at $E_{0}(\mathbf{k})=0$ isolated from two dispersive bands at $E_{\nu}(\mathbf{k})=\nu E(\mathbf{k})$ ($\nu=\pm$), with $E(\mathbf{k})=\sqrt{\Delta^{2}+2|\xi(\mathbf{k})|^{2}}$ and
$\xi(\mathbf{k})=t_{0}(1+e^{i\mathbf{k}\cdot\mathbf{a}_{1}}+e^{i\mathbf{k}\cdot\mathbf{a}_{2}})$ \cite{xu17,ocampo17}. In the hexagonal first Brillouin zone, the flat band approaches the two dispersive bands at $\mathbf{K}=(\frac{\sqrt{3}}{2},-\frac{1}{2})\frac{4\pi}{3a}$ and $\mathbf{K}^{\prime}=(\frac{\sqrt{3}}{2},\frac{1}{2})\frac{4\pi}{3a}$. Referring to the valley-contrasting optical excitations in inversion-broken graphene and transition metal dichalcogenide monolayers \cite{yao08,xiao12,cao12}, it is natural to ask if the present model also shows valley-contrasting interband excitations.

To trigger the interband transitions in the neutral fermionic cold atoms confined in the optical lattice, we exert a weak harmonic stimulus by sinusoidally shaking the lattice, which amounts to periodically modulating the phases of the laser beams \cite{tokuno11,wu15,anderson19,jotzu14,tran17,asteria19}. We consider a resonant circular driving to explore possible valley-contrasting excitations \cite{jotzu14,tran17,asteria19}. The strength of the vertical interband transition at $\mathbf{k}$ is proportional to (see Appendix F)
\begin{equation}
\langle\psi_{f}(\mathbf{k})|[\frac{\partial h(\mathbf{k})}{\partial k_{x}}+i\eta\frac{\partial h(\mathbf{k})}{\partial k_{y}}]
|\psi_{i}(\mathbf{k})\rangle,
\end{equation}
where $|\psi_{i}(\mathbf{k})\rangle$ and $|\psi_{f}(\mathbf{k})\rangle$ are the initial and final states of the interband transition at $\mathbf{k}$, $\eta=\pm$ marks the chirality of the circularly polarized stimulus. We consider $1/3$-filled and $2/3$-filled bands, for which the band edges situate at the two $\mathbf{K}$ points of the Brillouin zone. For circular modulations resonant with the transitions between the highest occupied states and the lowest empty states, that is for $\hbar\omega_{0}\simeq|\Delta|$ with $\omega_{0}$ being the angular frequency of the shaking, only states close to the $\mathbf{K}$ points contribute. We thus concentrate on these states and introduce the relative momenta $\mathbf{q}=\mathbf{k}-\mathbf{K}_{\tau}$, where $\tau=\pm$, $\mathbf{K}_{+}=\mathbf{K}^{\prime}$, and $\mathbf{K}_{-}=\mathbf{K}$.

For $1/3$-filled bands, $\psi_{i}=\psi_{-}$ and $\psi_{f}=\psi_{0}$. For $2/3$-filled bands, $\psi_{i}=\psi_{0}$ and $\psi_{f}=\psi_{+}$. For both cases and to the leading order of $\mathbf{q}$, the matrix element defined by Eq.(6) turns out to be (see Appendix F)
\begin{equation}
\frac{\sqrt{3}a}{2}|t_{0}|[\text{sgn}(\Delta)-\eta\tau]\frac{q_{x}+i\eta q_{y}}{q},
\end{equation}
where the sign function $\text{sgn}(x)=x/|x|$ for $x\neq0$. For fixed $\Delta$, the transition is nonvanishing only for the circular driving satisfying $\eta\tau$$=$$-\text{sgn}(\Delta)$. The chirality $\eta$ changes sign as the valley index $\tau$ changes sign. Therefore, the interband transitions involving the flat band are valley contrasting. It is interesting to explore the valley-contrasting physics in this displaced dice model with an isolated flat band, in analogy to that in monolayer transition metal dichalcogenides and gapped graphene \cite{yao08,xiao12,cao12}.

\section{Summary and discussion}

We have illustrated a general scheme of assembling multilayer optical lattices of cold atoms in a layer-by-layer manner. This allows us to simulate the properties of novel multilayer systems both from and beyond the condensed matter systems. As an example, we propose the optical devices to synthesize the displaced dice lattice that may realize the pure and generalized dice models, which have not been realized in experiment despite intensive theoretical studies. For a symmetrically biased dice model, we find valley-contrasting interband transitions associated with a flat band. Hopefully, because all the involved optical elements are standard, experimental realization of the proposal will come true soon.

The optical device in Figure 3 for the displaced dice lattice can easily be adjusted to apply to other novel multilayer optical lattices. Firstly, by controlling the orientations of the MMIs and so the propagation directions of the output laser beams, we may realize multilayer optical lattices with other lattice geometries, such as the square lattice, the Kagome lattice, and so on. In addition, by combining more than one sets of MMIs, we may realize multilayer optical lattices in which the constituent layers have hybrid lattice geometries (e.g., a multilayer of alternate square lattices and triangular lattices) \cite{grab16}. By removing the triad of BS$_{5}$, P$_{2}$ and BM$_{1}$ from Fig.3(b), the ensuing reduced MMIs may be used to create bilayer optical lattices. By inserting more triads of beamsplitters, plates, and bimirrors to the MMIs of Fig.3(b), on the other hand, we can make optical lattices with more than $3$ layers. Secondly, by controlling the phase parameters of the laser beams, we may achieve relative slipping between different layers along the layer plane. For the displaced dice lattice, this can be seen from the discussions in Sec.III and Eq.(F20) of Appendix F. Thirdly, which is of great current interest, we may twist the layers of the multilayer optical lattice by a continuously tunable angle. Referring to Fig.3 for the displaced dice lattice, by rotating the intersection edge of the bimirror in the $xy$ plane off the direction perpendicular to the propagation direction of the incoming laser beam, we may twist the reflected laser beam by an in-plane angle (see Appendix C). In this manner, we may twist a layer by rotating the propagation directions of the three laser beams for this layer to the same amount in the $xy$ plane, and consequently control the relative orientations between consecutive layers. Here, the twisting between consecutive layers may be implemented independently and up to a large twisting angle, which seem to be difficult in conventional solid state multilayer systems such as the bilayer graphene and thin films of the transition metal dichalcogenides.

Exploration of the above extensions to other multilayer optical lattices, and incorporation of many-body correlations, constitute highly intriguing future studies.


\begin{appendix}

\section{Transformation of the Gaussian beam by the cylindrical Keplerian telescope}

The beam shaper in our scheme is a cylindrical Keplerian telescope. In this section, we provide explicit mathematical definitions for the parameters characterizing the Gaussian beam input to and output from the cylindrical Keplerian telescope. This is done by summarizing well-known formulae for the transformations of the Gaussian beams by conventional (paraxial) optical elements \cite{sveltobook,siegmanbook}.

We refer to Figure 1(b) in the main text for the setup. The apertures of L$_{1}$ and L$_{2}$ are commonly much larger than the diameter of the input Gaussian laser beam, which is determined by the output aperture of the laser device and typically in the order of 1 mm \cite{sveltobook}. Therefore, the diffraction of the laser beam by the lenses L$_{1}$ and L$_{2}$ should be minor effects and is ignored in the analysis of this work.

We consider the propagation of the laser beam parallel to the $x$ axis, according to Fig. 1(b) of the main text. Before entering the beam shaper, the complex amplitude of the electric field of standard TEM$_{00}$ mode Gaussian laser beam can be written as \cite{sveltobook}
\begin{equation}
E(x,y,z)=\frac{A_{0}}{\rho(x)}\text{exp}\{-ik[x+\frac{y^2+z^2}{2q_{1}(x)}]+i\phi(x)\},
\end{equation}
where $k=2\pi/\lambda$ is the wave number, $A_{0}/\rho(x)$ is the amplitude of the electric field along the $x$ axis ($y=z=0$), $\rho(x)=\rho_{0}\sqrt{1+(\frac{\lambda x}{\pi\rho_{0}^{2}})^2}$ is the $x$-coordinate dependent radius of the disklike laser beam profile. $\rho_{0}$ is the waist radius at the waist plane ($x_{0}=0$ here) of the input Gaussian beam. Usually, $\rho_{0}$ is much larger than the wavelength of the laser (i.e., $\rho_{0}\gg\lambda$), so that $\rho(x)$ increases slowly as $x$ departs from the waist of the beam. A parameter characterizing the increase of $\rho(x)$ is the Rayleigh range $z_{R}$, which is the distance from the waist at which $\rho(x_{0}+z_{R})=\sqrt{2}\rho(x_{0})$. From the above definition of $\rho(x)$, we see that
\begin{equation}
z_{R}=\pi\rho_{0}^{2}/\lambda.
\end{equation}
A section of the beam centering at the waist and with a length 2$z_{R}$ is usually taken as the range over which the expansion of the Gaussian beam is small and negligible. In terms of the Rayleigh range, $\rho(x)=\rho_{0}\sqrt{1+(\frac{x}{z_{R}})^2}$. The phase factor $\phi(x)=\text{arctan}(\frac{\lambda x}{\pi\rho_{0}^{2}})=\text{arctan}(\frac{x}{z_{R}})$. Twice of the Rayleigh range (i.e., $2z_{R}$), or sometimes the Rayleigh range $z_{R}$ itself, is also called the confocal parameter of the Gaussian beam \cite{siegmanbook}.

Being neither plane wave nor spherical wave, the Gaussian beam is characterized by the $q_{1}(x)$ parameter defined as \cite{sveltobook}
\begin{equation}
\frac{1}{q_{1}(x)}=\frac{1}{R(x)}-i\frac{\lambda}{\pi\rho(x)^{2}}.
\end{equation}
$R(x)=x[1+(\frac{z_{R}}{x})^{2}]$ is the $x$-dependent curvature radius of the equiphase surface. The waist of the Gaussian beam at the $x=x_{0}=0$ plane is special and has an infinite curvature, namely $R(0)=\infty$ and $q_{1}(0)=iz_{R}$. The beam profile determined by Eq.(A1) on the $x$ plane is related to the beam profile on the $x=0$ waist plane, through a free propagation of $x$ distance in the free space. Namely, the field amplitude on the $x=0$ plane,
\begin{equation}
E(0,y,z)=\frac{A_{0}}{\rho_{0}}\exp\{-ik\frac{y^2+z^2}{2q_{1}(0)}\}=\frac{A_{0}}{\rho_{0}}\exp\{-\frac{y^2+z^2}{\rho_{0}^{2}}\},
\end{equation}
is transformed to
\begin{equation}
E(x,y,z)e^{ikx}=\frac{A_{0}}{\rho_{0}}\frac{1}{A+[B/q_{1}(0)]}\exp\{-ik\frac{y^2+z^2}{2q_{1}(x)}\},
\end{equation}
where
\begin{equation}
q_{1}(x)=\frac{Aq_{1}(0)+B}{Cq_{1}(0)+D}.
\end{equation}
The four coefficients are the components of the ABCD matrix for a free-space propagation of distance $x$, which is known to be \cite{sveltobook,siegmanbook}
\begin{equation}
\begin{pmatrix} A & B \\ C & D \end{pmatrix}=\begin{pmatrix} 1 & x \\ 0 & 1 \end{pmatrix}.
\end{equation}
Substituting the above coefficients into Eqs.(A5) and (A6), and defining the phase factor $\phi(x)=\arctan(\frac{x}{z_{R}})$, we reproduce Eq.(A1).

The free space as an optical medium is isotropic and transforms the Gaussian beam identically in both the $y$ direction and the $z$ direction. In an anisotropic medium (i.e., an astigmatic medium), however, the transformation matrix may be different along different directions. In this case, we treat the two directions independently and get two independent factors of the electric field amplitude, which are then multiplied together to give the overall electric field amplitude of the transformed laser beam \cite{siegmanbook}.

The cylindrical Keplerian telescope defined by Fig.1(b) of the main text, the beam shaper in our proposal, happens to be an astigmatic optical device. It reshapes only the $z$ axis amplitude distribution of the input Gaussian beam. The transformation of the $y$ axis amplitude distribution of the input Gaussian beam is just like the transformation in the free space. This anisotropic reshaping is conveniently represented by introducing two new $q$ parameters, $q_{2}$ and $q_{3}$, so that the electric field amplitude of the output laser beam becomes
\begin{equation}
\frac{A^{\prime}_{0}}{\sqrt{\rho_{2}(x)\rho_{3}(x)}}\text{exp}\{-ik[x+\frac{y^2}{2q_{2}(x)}+\frac{z^2}{2q_{3}(x)}]+i\phi_{23}(x)\},
\end{equation}
If the loss of the laser in passing through the telescope is negligible, we may set $A^{\prime}_{0}=A_{0}$. $\rho_{2}(x)$ and $\rho_{3}(x)$ are associated to $q_{2}$ and $q_{3}$, just as $\rho(x)$ is related to $q_{1}$ according to Eq.(A3).
Before passing the beam shaper, we have $q_{2}(x)=q_{3}(x)=q_{1}(x)$. After passing the beam shaper, $q_{2}(x)=q_{1}(x)$ still holds. However, $q_{3}(x)$ is different from $q_{1}(x)$ and transforms from $q_{1}(0)$ according to the overall ABCD matrix of the telescope system in the $z$ direction. The phase factor $\phi_{23}(x)$ is the average of the phase factors related to the transformations in the $y$ and the $z$ directions,
\begin{equation}
\phi_{23}(x)=\frac{\phi_{2}(x)+\phi_{3}(x)}{2}.
\end{equation}
$\phi_{2}(x)=\arctan(\frac{\lambda x}{\pi\rho_{0}^{2}})$. $\phi_{3}(x)$ is related to the components of the overall ABCD matrix of the cylindrical Keplerian telescope in the $z$ direction through \cite{siegmanbook}
\begin{equation}
e^{-i\phi_{3}(x)}=\frac{A+[B/q_{3}(0)]}{|A+[B/q_{3}(0)]|},
\end{equation}
where $q_{3}(0)=q_{2}(0)=q_{1}(0)=iz_{R}$.

The overall ABCD matrix of a composite optical system is the matrix multiplication of the ABCD matrices of the optical elements therein in sequence. We consider an input Gaussian laser beam with the waist at the $x=0$ plane at a distance $l_{1}$ to the left of L$_{1}$, and consider the output laser beam at a distance $l_{2}$ to the right of L$_{2}$. The telescope system is equivalent to a composition of five optical elements: a free space of length $l_{1}$, L$_{1}$ of focal length $f_{1}$, a free space of length $f_{1}+f_{2}\equiv L$, L$_{2}$ of focal length $f_{2}$, and a free space of length $l_{2}$.
The ABCD matrix for a thin lens whose focal length is $f$ is known as \cite{sveltobook,siegmanbook}
\begin{equation}
\begin{pmatrix} 1 & 0 \\ -\frac{1}{f} & 1 \end{pmatrix}.
\end{equation}
Together with the ABCD matrix for a free-space propagation for a length of $l$, defined by setting $x=l$ in Eq.(A7), we get the overall ABCD matrix for the $z$-axis transformation of the cylindrical Keplerian telescope by multiplying the five ABCD matrices together in sequence as
\begin{equation}
\begin{pmatrix} A & B \\ C & D \end{pmatrix}=\begin{pmatrix} -\frac{1}{M} & L-\frac{l_{1}}{M}-Ml_{2} \\ 0 & -M \end{pmatrix},
\end{equation}
where $M=f_{1}/f_{2}$ is the magnification of the telescope in the $z$ direction. The $q$ parameter at the waist of the incoming laser beam is $q_{1}(0)=i\pi\rho^{2}_{0}/\lambda=iz_{R}$. After passing through the cylindrical Keplerian telescope, the $q$ parameter in the $z$ direction becomes
\begin{equation}
q_{3}(l_{2})=\frac{Aq_{1}(0)+B}{Cq_{1}(0)+D}=\frac{q_{1}(0)+l_{1}+M^{2}l_{2}-ML}{M^{2}}.
\end{equation}
By writing
\begin{equation}
\frac{1}{q_{3}(l_{2})}=\frac{1}{R_{3}(l_{2})}-i\frac{\lambda}{\pi\rho_{3}^{2}(l_{2})},
\end{equation}
and comparing the two sides of Eq.(A14), we get
\begin{equation}
R_{3}(l_{2})=\frac{A^{2}z_{R}^{2}+B^{2}}{BD},
\end{equation}
\begin{equation}
\frac{\pi\rho_{3}^{2}(l_{2})}{\lambda}=\frac{A^{2}z_{R}^{2}+B^{2}}{ADz_{R}}.
\end{equation}
For fixed $l_{1}$, the waist for the $z$-axis amplitude distribution of the output laser beam is determined by $R_{3}(\tilde{l}_{2})=\infty$, which gives
\begin{equation}
\tilde{l}_{2}=\frac{L-\frac{l_{1}}{M}}{M}.
\end{equation}
Substituting into the expression for $\rho_{3}(l_{2})$, we get the waist size $\rho_{0z}$ for the amplitude distribution along the $z$ axis for the output laser beam as
\begin{equation}
\rho_{0z}=\rho_{3}(\tilde{l}_{2})=\frac{\rho(0)}{M}=\frac{\rho_{0}}{M}.
\end{equation}
Taking the two waist size parameters $\rho_{0}$ and $\rho_{0z}$ as references, the laser beam is compressed by $M$ times in the $z$ direction. Therefore, the cylindrical Keplerian telescope transforms the disklike beam profile of the input Gaussian beam to the straight filamentary beam profile of the output beam. Equivalently, the rod-like input Gaussian laser beam becomes the slab-like output laser beam.

Note that, the waist of the amplitude distribution along the $y$ direction is not changed by the telescope system and is still located at the plane $l_{1}$ to the left of L$_{1}$. We may want to change the waist position for the amplitude distribution of the output laser beam along $y$ to the same position as that along $z$. Or alternatively, we may want to make the output laser beam have a uniform distribution along the $y$ direction. This objective can be realized by replacing the simple cylindrical telescope used in Fig.1(b) of the main text by more sophisticated optical systems \cite{daniels03,homburg07}. However, because of the excellent unidirectionality of laser, one may take as a very good approximation to ignore the departure from the unidirectional propagation along $y$. This should be allowed in particular in the laboratory where the optical distances are of limited range. Therefore, we adhere in this work to the simple cylindrical telescope shown in Fig.1(b) of the main text. The waist size parameters $\rho_{0}$ and $\rho_{0z}$ are taken to measure the width and thickness of the beam profile of the output laser beam.


Besides technical limitations, and to facilitate the accurate control over the interlayer distances of the multilayer optical lattices, the reduced Rayleigh range for the amplitude distribution along the $z$ axis also requires us to relax the restriction in the thickness of the slab-like output laser beam. Because as $\rho_{0z}$ decreases, the far-field divergence angle of the beam along $z$ increases as $1/\rho_{0z}$ \cite{sveltobook}, the unidirectionality of the laser beam is weakened. As a result, the spatial extent, over which the thickness in the $z$-axis amplitude distribution of the laser beam could be regarded approximately constant, decreases monotonically with $\rho_{0z}$. Quantitatively, the range (along $x$, centering at the waist) over which the thickness of the laser along the $z$ direction may be regarded as constant is approximately twice the Rayleigh range, $2z_{R}=2\pi\rho_{0z}^{2}/\lambda$. If $\rho_{0z}\simeq\lambda$ is achieved, the spatial region over which a uniform 2D optical lattice may be defined hosts only about 100 unit cells of the optical lattice. For $\rho_{0z}\simeq5\lambda$, on the other hand, the approximately uniform region of the resulting multilayer optical lattice may have more than $6\times10^{4}$ unit cells. This should be enough for most quantum simulations.


Finally, we comment on the position of the $z$-axis waist of the compressed Gaussian beam. According to Eq.(A17), we have $\tilde{l}_{2}\simeq f_{2}$ for $l_{1}\simeq0$ and $M\gg1$. In the layer-by-layer scheme for the multilayer optical lattices, we wish to have the $z$-axis waist positions of the various beams approximately coincide with each other at the center of the designated location for the optical lattice. To be able to insert the optical elements in Figs. 3(a) and 3(b) in between L$_{2}$ of the cylindrical telescope and the location for the optical lattice, $\tilde{l}_{2}$ and thus $f_{2}$ should be large enough. For practical reasons, $f_{1}$ cannot be too large. Therefore, the magnification $M$ cannot be large. This means that, to achieve the desired compression ratio along the $z$ axis (i.e., about 200), we have to cascade two or more telescopes. In the last telescope, we can choose its $f_{2}$ larger than the preceding telescopes and a smaller magnification factor. For example, we can cascade two cylindrical telescopes of magnification 10 and a third cylindrical telescope with a magnification of 2 to achieve a total magnification of 200. In the last telescope, we may choose $f_{2}\simeq10$ cm, which gives $\tilde{l}_{2}\simeq 15$ cm for $l_{1}\simeq0$. By minimizing the sizes of the optical elements in the device, we should be able to bring the $z$-axis waist positions of the various thin-line beams to the center of the designated spatial region [e.g., within the circle of Fig.3(a)].
Besides increasing $f_{2}$, another method is to add a proper cylindrical lens to each MMI and refocus the outgoing beams, so that their $z$-axis waists are at or close to the center of the optical lattice.
Still another, more radical, strategy is to use diffractionless laser beams (in particular, the Bessel beams \cite{durnin87,herman91,mcgloin05}) with a thin-line beam profile. This strategy not only resolves the issue of small $\tilde{l}_{2}$ but may also enlarge the area of the ensuing multilayer optical lattice. It however requires more analysis as regards how to achieve this novel non-diffracting thin-line beam with sufficient intensity, which we leave to future studies.

\section{The 2D limit of an individual layer}

In terms of the line beam output from the beam shaper introduced in Fig.1(b) of the main text, we may form a quasi-2D optical lattice with reduced thickness characterized by $\rho_{0z}$. In applying to our layer-by-layer assembling of multilayer optical lattices, it is crucial to know the upper limit of $\rho_{0z}$ required for bringing the thickness of the resulting optical lattice down to the 2D limit.

Physically, if the input Gaussian laser beam is not compressed in the $z$ axis, each 2D site of the ensuing 2D optical lattice is in fact a quasi one-dimensional (q-1D) tube \cite{ozawa17}, for which the spectrum is a quasi-continuous 1D spectrum. As we compress the $z$-axis breadth (i.e., $\rho_{0z}$) of the laser beam, the q-1D spectrum becomes more and more discrete. Let us label the bound states in each site (i.e., potential well) of the 2D optical lattice by $\epsilon_{0}$, $\epsilon_{1}$, $\epsilon_{2}$, $\ldots$, in an order of increasing with the subscript. Then the site should be considered as a real zero-dimensional point rather than a q-1D tube when $\epsilon_{1}-\epsilon_{0}>E_{R}=h^{2}/(2m\lambda^{2})$. $E_{R}$ is the recoil energy of the optical lattice. If this condition is satisfied, only the lowest bound state of each quantum well is occupied. A tight-binding model for the cold atoms loaded to the optical lattice may be constructed by retaining only the lowest bound state for each site of the 2D lattice.

To be concrete, let us consider the optical dipole potential defined in Eqs.(1)-(4) of the main text for a single layer of the displaced dice lattice, which is a triangular lattice. By redefining the origin of coordinates, we may set the phase factors to zero. The optical dipole potential is written as
\begin{equation}
U(x,y,z)=-V_{0}\{1+4\cos\frac{bx}{2}[\cos\frac{bx}{2}+\cos\frac{\sqrt{3}by}{2}]\}g^{2}(z),
\end{equation}
where $b=4\pi/(\sqrt{3}a)$ is the magnitude of the reciprocal lattice vector, $g(z)=\exp(-z^{2}/\rho_{0z}^{2})$. We consider a red-detuned laser, so that \cite{grynberg01}
\begin{equation}
V_{0}=\frac{\epsilon_{0}}{4}\alpha_{0}'E_{0}^{2}>0.
\end{equation}
$(x,y,z)=(0,0,0)$ is clearly the bottom of one potential well. For simplicity, we focus on this potential well. Since we are interested in the 2D limit of the optical potential, we assume sufficiently deep potential wells and that the low-lying bound states are confined close to the bottom of the potential well. According to this assumption, we expand $U(x,y,z)$ to the leading order polynomials of $x$, $y$, and $z$ close to $(x,y,z)=(0,0,0)$. We get
\begin{eqnarray}
U(x,y,z)&\simeq&-9V_{0}+\frac{8\pi^{2}V_{0}}{a^{2}}(x^{2}+y^{2})+\frac{18V_{0}}{\rho_{0z}^{2}}z^{2}  \notag \\
&=&-9V_{0}+U(x)+U(y)+U(z).
\end{eqnarray}
The above potential energy term plus the kinetic energy term of the cold atom define an anisotropic 3D harmonic oscillator. We choose $\{\hat{H}_{x},\hat{H}_{y},\hat{H}_{z}\}$ as the complete set of commuting observables, where
\begin{equation}
\hat{H}_{\alpha}=\frac{p_{\alpha}^{2}}{2m}+U(\alpha),
\end{equation}
$\alpha=x,y,z$, $m$ is the mass of the cold atoms in the optical lattice.
The eigen-spectrum of the above anisotropic harmonic oscillator is
\begin{equation}
E_{n_{x}n_{y}n_{z}}=-9V_{0}+(n_{x}+n_{y}+1)\hbar\omega_{xy}+(n_{z}+\frac{1}{2})\hbar\omega_{z},
\end{equation}
where $n_{\alpha}$ ($\alpha=x,y,z$) are nonnegative integers quantifying the quantization of the energy spectrum. $\omega_{xy}=\frac{4\pi}{a}\sqrt{\frac{V_{0}}{m}}$ and $\omega_{z}=\frac{6}{\rho_{0z}}\sqrt{\frac{V_{0}}{m}}$ are separately the angular frequencies for the center-of-mass motion of the cold atom in the $xy$ plane and along the $z$ direction.

In order for the single layer to be in the 2D limit, the separation between the lowest energy bound state and the second lowest energy bound state should be larger than the recoil energy of the optical lattice. We therefore require $\hbar\omega_{xy}>E_{R}$ and $\hbar\omega_{z}>E_{R}$. For our purpose, considering the continuous compression of $\rho_{0z}$ from $\rho_{0}$, we originally have $\rho_{0z}\gg a$ and correspondingly $\omega_{z}/\omega_{xy}=\frac{3}{2\pi}\frac{a}{\rho_{0z}}\ll1$. We assume the $\hbar\omega_{xy}>E_{R}$ condition is always satisfied for all $\rho_{0z}$. As we compress $\rho_{0z}$ continuously, the 2D limit is attained when $\hbar\omega_{z}>E_{R}$. Suppose $\hbar\omega_{xy}=\gamma E_{R}$ ($\gamma\gg1$), the condition $\hbar\omega_{z}>E_{R}$ amounts to
\begin{equation}
\rho_{0z}<\frac{3\gamma}{2\pi}a.
\end{equation}
For our triangular lattice, $a=\frac{2}{3}\lambda$. For $\gamma>5\pi\simeq15.71$, $\rho_{0z}=5\lambda$ satisfies the above constraint and the 2D limit is attained. For $^{40}$K cold atoms, red-detuned laser with a wavelength 1064 nm or 1030 nm were used in experiments \cite{jordens08,tarruell12,greif13,schneider08,hackermuller10}. Correspondingly, $\rho_{0z}=5\lambda$ amounts to $\rho_{0z}\approx5$ $\mu$m.

From the definition of the angular frequency, $\hbar\omega_{xy}=\gamma E_{R}>15.71E_{R}$ is equivalent to $V_{0}=\frac{\gamma^{2}}{18}E_{R}>13.71E_{R}$. This is an intermediate value for the lattice depth. However, it is $9V_{0}$ from the minima to the maxima of the optical dipole potential defined by Eq.(B1). Therefore, $9V_{0}$ is the overall lattice depth, which should satisfy $9V_{0}>123.39E_{R}$. As a result, we need to work with a very deep optical lattice to attain the 2D limit for an individual triangular optical lattice. Luckily, this depth is within the reach of experiments \cite{grynberg01,raithel97}. Clearly, for even deeper optical dipole potential, $\rho_{0z}$ larger than 5$\lambda$ may also lead to the 2D limit. Conversely, if we compress the Gaussian beam thinner, with $\rho_{0z}<5\lambda$, a shallower optical potential may also attain the 2D limit. For example, if we consider $\rho_{0z}=3\lambda$, $9V_{0}>44.5E_{R}$ is enough to get the 2D limit for the triangular optical lattice.

Finally, we should emphasize that the analysis in this section is based on the optical dipole potential of a single layer. In a multilayer optical lattice containing several contiguous layers, such as the displaced dice lattice consisting of three consecutive layers, the superposition of the optical potentials for the various layers may possibly change the actual depth of the resulting multilayer optical dipole potential. From Figure 2 of the main text for the lattice geometry  and the expressions of the optical dipole potentials of the various layers, this superposition tends to reduce the actual depth of the optical potential for the displaced dice lattice. From the analysis carried out later in Appendix E, the superposition does not change $\omega_{z}$ in the leading order approximation. Therefore, the above criterion for the 2D limit is still applicable. However, the overall trilayer optical dipole potential confining the cold atoms in the displaced dice lattice is actually shallower than the potential of a single layer. The shorter the distances between the three layers, the shallower the trilayer optical dipole potential becomes. For other multilayer optical lattices synthesized through the layer-by-layer approach, the depth of the overall multilayer optical dipole potential may also be larger than the depth of a single layer. At the same time, the critical depth of the lattice for attaining the 2D limit may also change quantitatively.

\section{Reflection of a ray by the Fresnel bimirror with about 90$^{\circ}$ intersection angle}

In what follows, we call the special Fresnel bimirror with a 90$^{\circ}$ intersection angle as the right-angle bimirror. In Fig. 3 of the main text, the reflection plane of the ray of light is assumed to be perpendicular to the intersection edge of the right-angle bimirror. In this case, the retroreflected light propagates along the direction which is exactly opposite to the incident ray of light. This is always the case once the intersection edge of the right-angle bimirror is perpendicular to the incident ray of light, because by default the intersection edge is always perpendicular to the normal of the two mirrors. The lateral shift (i.e., the distance) between the incident and the retroreflected rays of light depends on the incident angle and the distance from the point of incidence to the intersection edge. As the orientation of the intersection edge has an additional $2\pi$ rotational freedom in the plane perpendicular to the incident ray of light, the lateral shift of the retroreflected ray can also occur at any direction away from the incident ray of light.

By twisting the intersection edge of the right-angle bimirror away from the plane perpendicular to the incident ray of light by a small angle $\theta$, the reflection plane of the incident ray of light is also slightly deviated from perpendicular to the intersection edge. In this case, we may twist the propagation direction of the retroreflected ray away from exactly the reverse direction of the incident ray. Here, the intersection edge at an angle $\theta$ to the plane perpendicular to the incident ray of light also has a $2\pi$ rotational freedom, which may be characterized by an angle $\varphi$. As regards the retroreflected ray of light, we expect there to be a corresponding freedom in the deviation of its propagation direction with respect to the direction exactly opposite to the incident ray.

Instead of plotting the full optical path, which is clumsy and not easy to show clearly the various interesting situations, we will express the reflection off the bimirror in terms of vector algebra. Because of the simplicity of the reflection by plane mirrors, exact analytical expressions for the outgoing reflected ray may be obtained.

For generality, we consider a Fresnel bimirror whose intersection angle $\alpha$ is close to but may be slightly different from 90$^{\circ}$. We notice that the reflected ray is uniquely determined by the position and orientation of the $\alpha$-angle bimirror, and the propagation direction and point of incidence of the incident ray. In terms of vector algebra, these conditions are contained in the two equations for the two mirror planes and the equation for the incident ray.

\begin{figure}\label{fig4} \centering
\includegraphics[width=8.6cm,height=7.92cm]{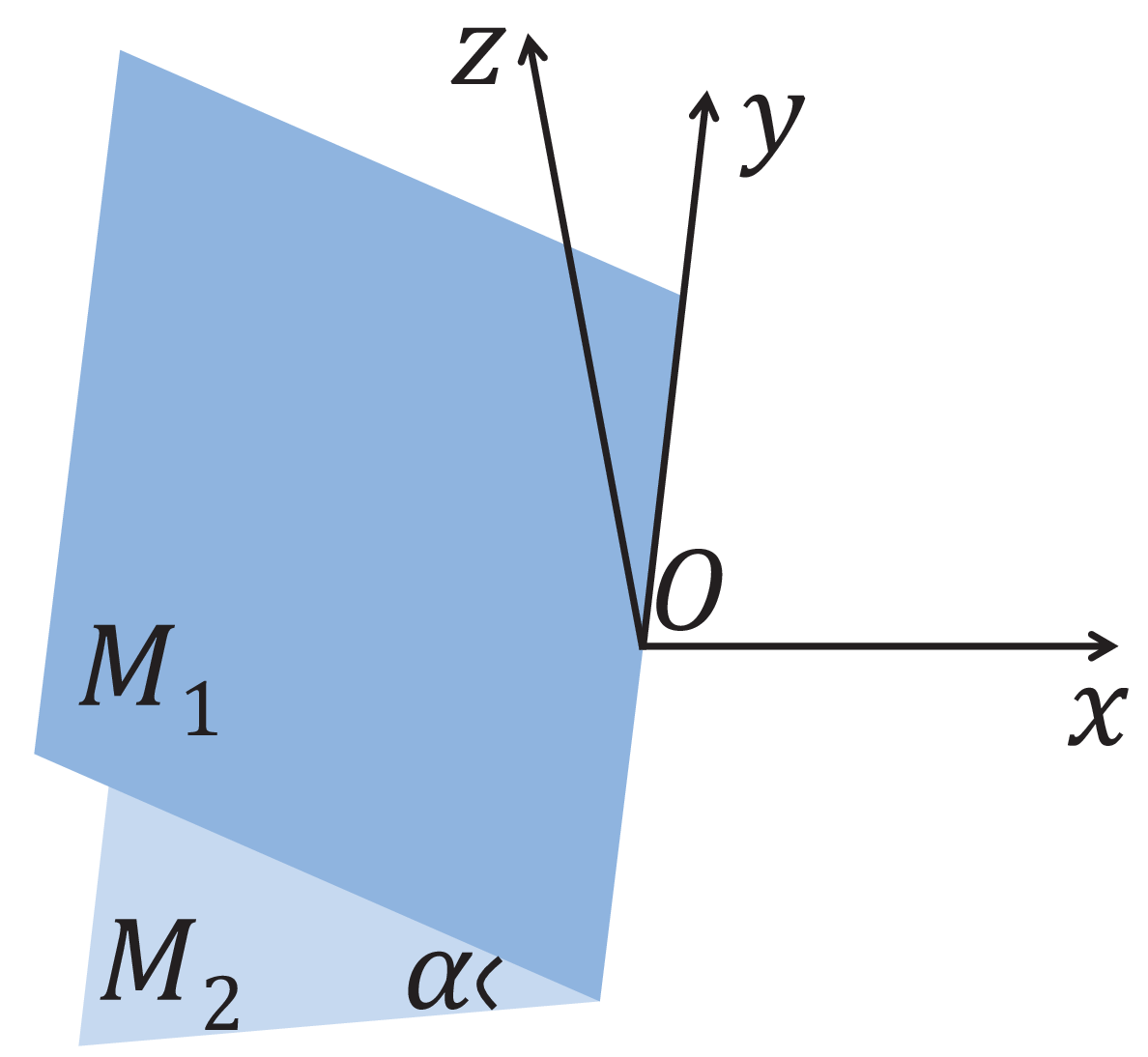}
\caption{The Fresnel bimirror with an intersection angle $\alpha$ between the two mirrors M$_{1}$ and M$_{2}$. The Cartesian coordinate is defined by taking the center of the intersection edge as the origin O, and the intersection edge as the $y$ axis. The $xy$ plane bisects the bimirror.}
\end{figure}

To proceed, we firstly define the coordinate system in reference to the bimirror shown in Figure 4 of this section. We consider a 3D Cartesian coordinate. We assume the intersection edge of the bimirror is along the $y$ axis, so that the planes perpendicular to the intersection edge are parallel to the $xz$ plane. An arbitrary perpendicular plane of the intersection edge crosses with the two mirrors at two lines, which join each other at a point of the intersection edge and subtend an angle of $\alpha$. We assume that the bimirror is divided equally by the $xy$ plane containing the intersection edge. We define the $x$ axis to run opposite to the open-mouth direction of the $\alpha$-angle bimirror, so that the incident ray propagates along a direction very close to the positive $x$ direction (i.e., close to $\hat{\mathbf{x}}$, the unit vector along the $+x$ direction). We set the origin $O$ of the coordinate, $(0,0,0)$, at the middle of the intersection edge. The two mirrors are assumed to be large enough so that the interesting incident rays are all reflected twice by the $\alpha$-angle bimirror before leaving it. Finally, we assume that the crossing point between the incident ray and the first mirror (e.g., M$_{1}$) is close to the intersection edge.

Under the above conditions, the equations for the two mirror planes are
\begin{equation}
\mathbf{r}\cdot\hat{\mathbf{n}}_{i}=0,
\end{equation}
where $i=1,2$ labels the two mirrors, $\hat{\mathbf{n}}_{1}=(\sin\frac{\alpha}{2},0,\cos\frac{\alpha}{2})$ and $\hat{\mathbf{n}}_{2}=(\sin\frac{\alpha}{2},0,-\cos\frac{\alpha}{2})$ are unit normal vectors for M$_{1}$ and M$_{2}$.

In the application of the main text, we fixed $\alpha=\pi/2$ and the incident ray propagates along the positive $x$ direction (i.e., $\hat{\mathbf{x}}$). In this case, the incident ray is exactly retroreflected, up to a lateral shift in the $z$ direction. Here, we assume the propagation direction of the incident ray may be slightly away from the $\hat{\mathbf{x}}$ direction by a small angle $\theta$. To characterize this direction, we introduce a spherical polar coordinate taking $\hat{\mathbf{x}}$ as the polar axis, $\theta$ as the polar angle, and define the azimuthal angle $\varphi$ as the angle with respect to the $\hat{\mathbf{y}}$ direction in the $yz$ plane. In this polar coordinate, the propagation of the incident ray is along the unit vector
\begin{equation}
\hat{\mathbf{u}}=(\cos\theta,\sin\theta\cos\varphi,\sin\theta\sin\varphi).
\end{equation}
Without losing generality, we assume that the incident ray shines at a point P on M$_{1}$ of a distance $d>0$ away from the intersection edge. The coordinate of P is written as
\begin{equation}
\mathbf{r}_{1}=(-d\cos\frac{\alpha}{2},y_{1},d\sin\frac{\alpha}{2}).
\end{equation}
For simplicity, we set $y_{1}=0$. The equation of the incident ray is thus
\begin{equation}
(\mathbf{r}-\mathbf{r}_{1})\times\hat{\mathbf{u}}=\mathbf{0}.
\end{equation}
The reflected ray leaves M$_{1}$ also at P. Assuming the unit vector along its propagation direction to be $\hat{\mathbf{u}}_{1}$, the law of reflection dictates that
\begin{equation}
\begin{cases}
\hat{\mathbf{u}}\cdot\hat{\mathbf{n}}_{1}=-\hat{\mathbf{u}}_{1}\cdot\hat{\mathbf{n}}_{1},  \\
\hat{\mathbf{u}}_{1}\cdot(\hat{\mathbf{u}}\times\hat{\mathbf{n}}_{1})=0,  \\
\end{cases}
\end{equation}
where the first equation ensures the reflection angle to be equal to the incident angle, and the second equation says that the reflected ray lies in the same plane containing the incident ray and the unit normal vector of M$_{1}$ passing P. Solution to Eq.(C5) gives
\begin{equation}
\hat{\mathbf{u}}_{1}=\hat{\mathbf{u}}-2(\hat{\mathbf{u}}\cdot\hat{\mathbf{n}}_{1})\hat{\mathbf{n}}_{1}.
\end{equation}
From $\hat{\mathbf{u}}\cdot\hat{\mathbf{u}}=\hat{\mathbf{n}}_{1}\cdot\hat{\mathbf{n}}_{1}=1$, we have $\hat{\mathbf{u}}_{1}\cdot\hat{\mathbf{u}}_{1}=1$. So $\hat{\mathbf{u}}_{1}$ is the unit vector along the propagation direction of the first reflected ray. The equation for the first reflected ray leaving M$_{1}$ at P is thus
\begin{equation}
(\mathbf{r}-\mathbf{r}_{1})\times\hat{\mathbf{u}}_{1}=\mathbf{0}.
\end{equation}

The coordinate $\mathbf{r}_{2}$ of the crossing point between the first reflected ray and M$_{2}$ is determined by solving the following simultaneous equations
\begin{equation}
\begin{cases}
(\mathbf{r}-\mathbf{r}_{1})\times\hat{\mathbf{u}}_{1}=\mathbf{0}, \\
\mathbf{r}\cdot\hat{\mathbf{n}}_{2}=0. \\
\end{cases}
\end{equation}
The solution $\mathbf{r}=\mathbf{r}_{2}=(x_{2},y_{2},z_{2})$ to the above equations is also the point at which the second reflected ray leaves M$_{2}$ and the whole bimirror. Similar to the first reflection, the unit vector $\hat{\mathbf{u}}_{2}$ along the propagation direction of the second reflected ray is determined by
\begin{equation}
\begin{cases}
\hat{\mathbf{u}}_{1}\cdot\hat{\mathbf{n}}_{2}=-\hat{\mathbf{u}}_{2}\cdot\hat{\mathbf{n}}_{2},  \\
\hat{\mathbf{u}}_{2}\cdot(\hat{\mathbf{u}}_{1}\times\hat{\mathbf{n}}_{2})=0,  \\
\end{cases}
\end{equation}
which gives
\begin{equation}
\hat{\mathbf{u}}_{2}=\hat{\mathbf{u}}_{1}-2(\hat{\mathbf{u}}_{1}\cdot\hat{\mathbf{n}}_{2})\hat{\mathbf{n}}_{2}.
\end{equation}
The equation for the second (i.e., final) reflected ray outgoing from the bimirror is
\begin{equation}
(\mathbf{r}-\mathbf{r}_{2})\times\hat{\mathbf{u}}_{2}=\mathbf{0}.
\end{equation}
The explicit expressions for $\mathbf{r}_{2}=(x_{2},y_{2},z_{2})$ and $\hat{\mathbf{u}}_{2}=(u_{2x},u_{2y},u_{2z})$ are
\begin{equation}
\begin{cases}
x_{2}=-\frac{\cos\theta\sin\frac{\alpha}{2}+\sin\theta\sin\varphi\cos\frac{\alpha}{2}} {\cos\theta\sin\frac{3\alpha}{2}+\sin\theta\sin\varphi\cos\frac{3\alpha}{2}}d\cos\frac{\alpha}{2},  \\
y_{2}=\frac{\sin\theta\cos\varphi\sin\alpha}{\cos\theta\sin\frac{3\alpha}{2}+\sin\theta\sin\varphi\cos\frac{3\alpha}{2}}d,  \\
z_{2}=-\frac{\cos\theta\sin\frac{\alpha}{2}+\sin\theta\sin\varphi\cos\frac{\alpha}{2}} {\cos\theta\sin\frac{3\alpha}{2}+\sin\theta\sin\varphi\cos\frac{3\alpha}{2}}d\sin\frac{\alpha}{2},  \\
\end{cases}
\end{equation}
and
\begin{equation}
\begin{cases}
u_{2x}=\cos\theta\cos(2\alpha)-\sin\theta\sin\varphi\sin(2\alpha),  \\
u_{2y}=\sin\theta\cos\varphi,  \\
u_{2z}=\cos\theta\sin(2\alpha)+\sin\theta\sin\varphi\cos(2\alpha).  \\
\end{cases}
\end{equation}

The lateral shift related to $\mathbf{r}_{2}-\mathbf{r}_{1}$ is linearly proportional to $d$ and could be tuned continuously. To characterize the change in the propagation direction, we study the direction cosines of $\hat{\mathbf{u}}_{2}$ with regards to the direction of $\hat{\mathbf{u}}$ and $\hat{\mathbf{u}}\times\hat{\mathbf{y}}$, which turn out to be
\begin{equation}
\hat{\mathbf{u}}_{2}\cdot\hat{\mathbf{u}}=-1+2[1-\sin^{2}\alpha(1-\sin^{2}\theta\cos^{2}\varphi)],
\end{equation}
and
\begin{equation}
\hat{\mathbf{u}}_{2}\cdot(\hat{\mathbf{u}}\times\hat{\mathbf{y}})=\sin(2\alpha)(\sin^{2}\theta\sin^{2}\varphi+\cos^{2}\theta).
\end{equation}
From these formulae we uncover the conditions for the following two especially interesting cases.

In the first case, which is the case we focused on in the main text, the reflected ray propagates in exactly the opposite direction to the incident ray. This is satisfied if
\begin{equation}
\hat{\mathbf{u}}_{2}\cdot\hat{\mathbf{u}}=-1,
\end{equation}
which leads to $\alpha=\pi/2$ and $\varphi=\pm\pi/2$. That is, the ray incident on an $\alpha=\pi/2$ right-angle bimirror propagates along a direction parallel to the $xz$ plane and close to $\hat{\mathbf{x}}$ (i.e., $\theta$ is small so that the incident ray undergoes two reflections off the bimirror).

In the second case, the reflected ray propagates almost opposite to the direction of the incident ray but deflected a little bit away from that opposite direction in the plane perpendicular to $\hat{\mathbf{u}}\times\hat{\mathbf{y}}$. This case is determined by
\begin{equation}
\begin{cases}
\hat{\mathbf{u}}_{2}\cdot\hat{\mathbf{u}}+1=\epsilon,  \\
\hat{\mathbf{u}}_{2}\cdot(\hat{\mathbf{u}}\times\hat{\mathbf{y}})=0, \\
\end{cases}
\end{equation}
where $0<\epsilon\ll1$. For $\theta\simeq0$ that we focus on, the only solution to the second equation of Eq.(C17) for $0<\alpha<\pi$ is $\alpha=\pi/2$. Substituting $\alpha=\pi/2$ into Eq.(C14), we have
\begin{equation}
\hat{\mathbf{u}}_{2}\cdot\hat{\mathbf{u}}+1=2\sin^{2}\theta\cos^{2}\varphi.
\end{equation}
Then the first equation of Eq.(C17) is satisfied once $\theta$ is small but nonzero and $\varphi\neq\pm\pi/2$. For our purpose of using the bimirror, it is enough to fix $\varphi=0$ or $\varphi=\pi$, so that $\hat{\mathbf{u}}$ is parallel to the $xy$ plane. In this case, suppose the rotation angle from $\hat{\mathbf{u}}$ to $\hat{\mathbf{u}}_{2}$ is $\pi+\tilde{\theta}$, we have
\begin{equation}
\hat{\mathbf{u}}_{2}\cdot\hat{\mathbf{u}}=\cos(\pi+\tilde{\theta})=-\cos\tilde{\theta}=2\sin^{2}\theta-1=-\cos(2\theta).
\end{equation}
Therefore, rotating the incident ray in the $xy$ plane by an angle of $\theta$ away from $\hat{\mathbf{x}}$ leads to a rotation of the reflected ray by an angle of 2$\theta$ in the $xy$ plane, compared to the direction of exact retroreflection. This rotation is also easy to see by substituting $\alpha=\pi/2$ and $\sin\varphi=0$ into the explicit expression for $\hat{\mathbf{u}}_{2}$ in Eq.(C13) and comparing it with the definition of $\hat{\mathbf{u}}$. The angle $\tilde{\theta}=2\theta$ may be called as the \emph{twist angle}. The twist angle $\tilde{\theta}$ enables a continuous control over the twisting angle between consecutive layers of the multilayer optical lattice constructed in the layer-by-layer manner.

For both of the two cases considered above, the right-angle bimirror may be replaced by an isosceles right-angle triangular prism (in brief, a right-angle prism) made of low-loss glass with a high refractive index. To see this, notice that the above analyses may be considered as corresponding to the two total internal reflections inside the interior of the right-angle prism. M$_{1}$ and M$_{2}$ of the right-angle bimirror are interpreted here as the two mutually perpendicular rectangular faces of the prism, and the incident ray enters the right-angle prism through the rectangular bevel face in between the two perpendicular faces. The actual incident ray is connected to the ray incident on the M$_{1}$ face at P by a refraction at the bevel face of the right-angle prism. The actual outgoing ray is connected to the ray leaving the M$_{2}$ face at $\mathbf{r}_{2}$ by another refraction at the bevel face of the right-angle prism. For the first case studied above, it is easy to see that the retroreflected ray still reverses the direction of the incident ray. In simple words, two parallel rays are still parallel rays upon refraction by a planar dielectric interface. For the second case, the actual twisting angle is changed from 2$\theta$ to 2$\theta'$. The actual twisting angle 2$\theta'$ is related to the twisting angle 2$\theta$ within the right-angle prism by Snell's law
\begin{equation}
\sin\theta'=n\sin\theta,
\end{equation}
where $n$ is the refractive index of the glass making up the right-angle prism. The refractive index for the air has been taken as 1.

The above analyses deal with a single ray of light. In considering the reflection of a laser beam, we regard it as an assembly of nearly parallel rays. The reflection of a line beam, whose long axis of the beam profile is within the $xy$ plane, by a right-angle bimirror or a right-angle prism follows the same rules explained above.

Finally, we notice that, rotating the incident ray (or, the incident line beam) in the $xy$ plane with respect to the $\hat{\mathbf{x}}$ direction is equivalent to fixing the incident ray to propagate along $\hat{\mathbf{x}}$ and rotating the intersection edge of the right-angle bimirror in the $xy$ plane by the same angle in the opposite sense (i.e., clockwise versus counterclockwise, and vice versa). In experimental implementations, it seems more convenient to control the bimirrors. Therefore, it is preferable to perform the rotations by acting upon the bimirrors, rather than upon the incident rays.

\section{Another optical device for the displaced dice lattice}

In application to the displaced dice lattice, the optical device in Fig. 3 of the main text firstly splits the single compressed Gaussian beam (i.e., the line beam) into three beams of equal intensities, and secondly splits each of the three beams again into three beams of equal or unequal intensities (but with the relative intensities the same for all three triplet sets) that are displaced from each other in the $z$ direction.

The optical device in Fig. 3 of the main text permits independent manipulation over all the nine laser beams. However, our interest in this work has been restricted to the displaced dice lattices where the lattice vectors of all the three layers are parallel to each other. In this `simple' configuration, we may split the single line beam into nine beams following the reverse sequence. Namely, we firstly split the line beam output from the beam shaper into three beams that have equal or unequal intensities and are displaced from each other along the $z$ direction but aligned in their projection in the $xy$ plane, and then split this compound beam into three compound beams of equal intensities. By guiding the three compound beams to the designated spot for the optical lattice, we again obtain a trilayer optical lattice. An MMI defined in Fig. 3(b) of the main text is certainly capable of implementing the first step. Figure 5 of this section has taken an alternative combination that shifts the laser beams in terms of tilted glass plates (i.e., according to the left of Fig. 1(c) of the main text).

\begin{figure}\label{fig5} \centering
\includegraphics[width=8.7cm,height=3.575cm]{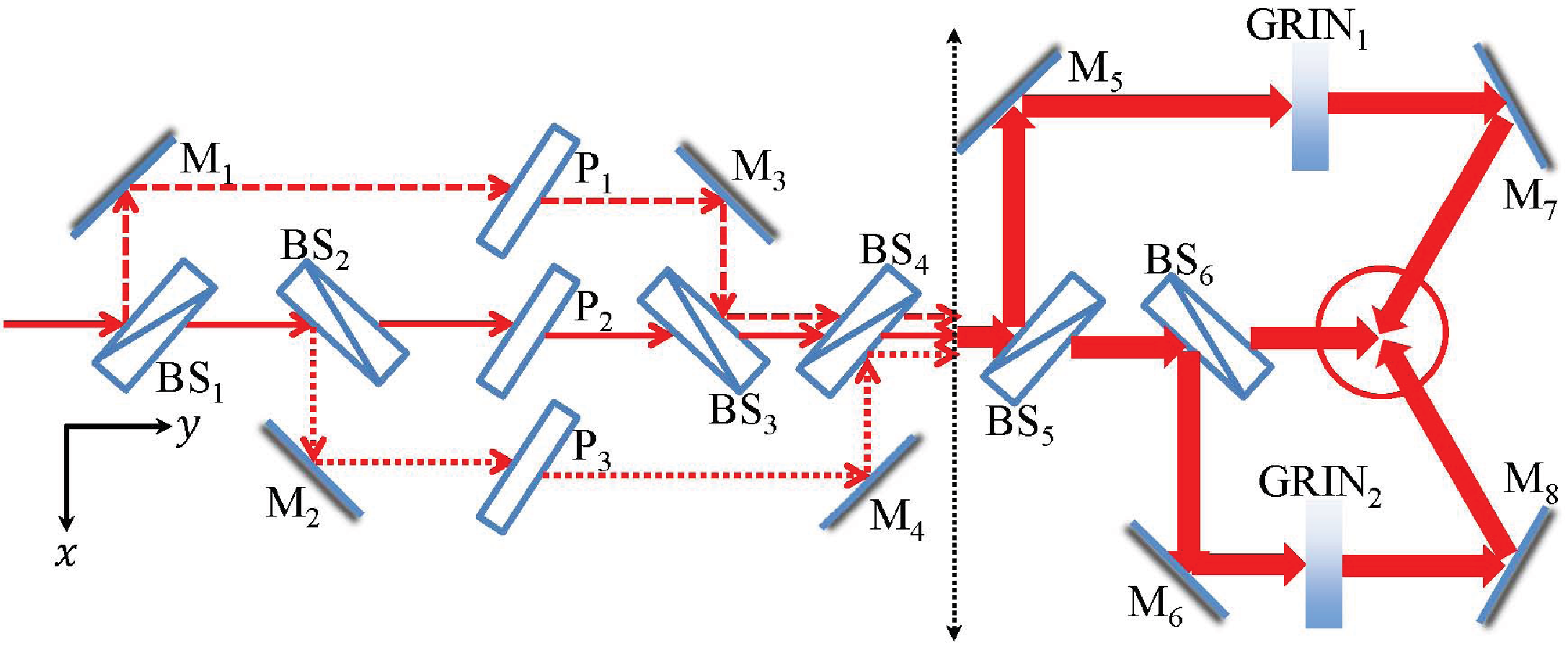}
\caption{Another optical device that transforms the compressed Gaussian beam into the displaced dice lattice. The biapiculate vertical dotted line separates the device into two parts. The left part splits the input line beam into three co-propagating beams displaced from each other along the $z$ direction. For clarity of illustration, the three beams have been \emph{drawn as} shifted in the $xy$ plane. The distance between the three beams (solid, dashed, and dotted) are controlled by the thicknesses and tilting angles of the three glass plates, P$_{1}$ through P$_{3}$. These three co-propagating beams output from the left part of the device are represented by a single thick arrowed line on the right part of the device. The right part of the device splits this compound beam again into three compound beams, propagating with intersection angle $\frac{2}{3}\pi$ between any two of the three compound beams, according to the three-beam protocol for the triangular optical lattice. GRIN is the shorthand for the gradient-index optical element. Another GRIN may be inserted to the compound beam on the right of BS$_{6}$.}
\end{figure}

One drawback of this alternative setup lies in the difficulty in adjusting the relative phase shifts of consecutive layers, which is crucial to lining up the three triangular lattices in the correct relative positions in the $xy$ plane. This is because, according to the analysis in the main text, the phase shifts should be implemented differently for the three beams inside a single compound beam. To implement such phase shifts, as shown on the right part of Fig. 5 of this section, we suggest to use two glass plates with a gradient in the refractive index. This kind of optical elements, although unconventional in comparison to the other optical elements employed (i.e., cylindrical thin lenses, beam splitters, mirrors, glass plates), are also standard optical elements in the subject of gradient-index (GRIN) optics \cite{gomez02}. According to the gradient-index optics, GRIN$_{1}$ and GRIN$_{2}$ in Fig.2 of this section belong to the GRIN medium with an axial gradient \cite{gomez02}. With finely tuned gradient in the refractive index, hopefully these GRIN optical elements are able to introduce the desired phase shifts. In even simpler multilayer optical lattices where no relative in-plane shifts between the consecutive layers are required, we do not need these GRIN optical elements and the scheme similar to that of Fig. 5 of this section may find more applications.

\section{Estimation of tight-binding parameters for the displaced dice lattice}

We show in this section that, by tuning the relative strength of the electric field vectors of the laser beams and the relative $z$-axis coordinates, we may realize the following family of tight-binding models for cold atoms on the displaced dice lattice
\begin{eqnarray}
\hat{H}&=&\sum\limits_{<i,j>,\sigma}(t_{ba}b^{\dagger}_{i\sigma}a_{j\sigma}+\text{H.c.})
+\sum\limits_{<i,j>,\sigma}(t_{bc}b^{\dagger}_{i\sigma}c_{j\sigma}+\text{H.c.})   \notag \\
&&+\sum\limits_{i,\sigma}(\varepsilon_{a}a^{\dagger}_{i\sigma}a_{i\sigma}
+\varepsilon_{b}b^{\dagger}_{i\sigma}b_{i\sigma}
+\varepsilon_{c}c^{\dagger}_{i\sigma}c_{i\sigma}).
\end{eqnarray}
By taking $\varepsilon_{b}$ as the reference energy, and defining $\varepsilon_{ab}=\varepsilon_{a}-\varepsilon_{b}$ and $\varepsilon_{cb}=\varepsilon_{c}-\varepsilon_{b}$, it is the same as Eq.(5) of the main text.
The parameters in the model follow the same definitions for Eq.(5) of the main text. To validate this model, the intralayer (i.e., intrasublattice) nearest-neighboring (NN) hopping amplitudes and the NN hopping amplitudes between the A layer and the C layer should be negligible as compared to the BA interlayer and BC interlayer NN hopping amplitudes. Put it another way, the retained BA interlayer NN hopping amplitudes and the BC interlayer NN hopping amplitudes should dominate among the six kinds of NN hopping amplitudes.

To justify the above approximation, we have to make a reasonable estimation over the magnitudes of these relevant NN hopping amplitudes (6 in total). We suppose all three layers of the displaced dice lattice have reached the 2D limit and the on-site interactions have been tuned to zero by the Feshbach resonance. We also assume that fermionic cold atoms occupy only the lowest-energy bound states of the potential wells. In this case, a reasonable order of magnitude estimation over the relevant hopping amplitudes follows by taking the ground state eigenvectors of the approximate harmonic models for each quantum well as the local Wannier orbitals and substitute them into the definition of the hopping amplitudes. A merit of this approach, despite less accurate, is that all hopping amplitudes may be evaluated in a fully analytical manner. Having an explicit analytical expression for the hopping amplitudes in terms of the parameters of the optical lattice greatly facilitates a qualitative understanding over the model parameters of the tight-binding model.

For the displaced dice lattice defined in the main text, the single-body Hamiltonian is
\begin{equation}
\hat{H}_{0}=-\frac{\hbar^{2}}{2m}\boldsymbol{\nabla}^{2}+V(x,y,z),
\end{equation}
where $m$ is the mass of the cold fermionic alkali-metal element such as $^{40}$K or $^{6}$Li. $V$ is the sum of the optical dipole potentials of all three layers
\begin{equation}
V(x,y,z)=U_{A}(x,y,z)+U_{B}(x,y,z)+U_{C}(x,y,z),
\end{equation}
where the subindices A, B, and C indicate the three triangular layers. Here, following the discussions of the main text, we do not include the interlayer interference terms in the optical potential. Analyses for their effects are deferred to future more quantitative calculations of the band structures. Taking the minimum of one potential well of the B layer as the origin of coordinate, $(x,y,z)=(0,0,0)$, we have
\begin{subequations}
\begin{equation}
U_{B}(x,y,z)=-V_{B}[1+4\cos\frac{bx}{2}(\cos\frac{bx}{2}+\cos\frac{\sqrt{3}by}{2})]g^{2}(z),
\end{equation}
\begin{equation}
U_{A}(x,y,z)=U_{B}(x,y,z)|_{x\rightarrow x-\frac{a}{\sqrt{3}},y\rightarrow y,z\rightarrow z-z_{A};V_{B}\rightarrow V_{A}},
\end{equation}
\begin{equation}
U_{C}(x,y,z)=U_{B}(x,y,z)|_{x\rightarrow x+\frac{a}{\sqrt{3}},y\rightarrow y,z\rightarrow z-z_{C};V_{B}\rightarrow V_{C}},
\end{equation}
\end{subequations}
where the Gaussian factor $g(z)=\exp{(-z^{2}/\rho_{0z}^{2})}$, the length of the reciprocal lattice vector $b=4\pi/(\sqrt{3}a)$.

Generally, the hopping amplitude between two NN sites (i.e., two NN minima of the optical potential) located at $\mathbf{r}_{0}=(x_{0},y_{0},z_{0})$ and $\mathbf{r}_{1}=(x_{1},y_{1},z_{1})$ is \cite{lewenstein07,lee09,tormabook}
\begin{equation}
t_{NN}(\mathbf{r}_{0}\mathbf{r}_{1})=\int d^{3}\mathbf{r}\psi^{\ast}_{0}(\mathbf{r}-\mathbf{r}_{0})\hat{H}_{0}\psi_{0}(\mathbf{r}-\mathbf{r}_{1}).
\end{equation}
In this definition, the layer (sublattice) index is implicitly included in the reference coordinates $\mathbf{r}_{0}$ and $\mathbf{r}_{1}$. $\psi_{0}(\mathbf{r}-\mathbf{r}_{0})$ is the Wannier function for the local orbital at $\mathbf{r}_{0}$.

For a qualitative estimation of the hopping amplitudes, we take $\psi_{0}$ as the ground-state wave function for a quantum well in the harmonic approximation. In Section II analyzing the condition for attaining the 2D limit, we focused on the optical dipole potential for a single layer. In the present displaced dice lattice containing three layers, the bottom of the potential well in one layer is also influenced by the optical potentials of the other two layers. Therefore, the approximate harmonic oscillator model for the quantum well in each layer should also be supplemented with the contribution from the other two layers. For the B-layer, we consider the potential well whose minima is at $(0,0,0)$. For the A-layer, we consider the potential well whose minima is at $(a/\sqrt{3},0,z_{A})$. For the C-layer, we consider the potential well whose minima is at $(-a/\sqrt{3},0,z_{C})$.

First, we consider the potential well of the A-layer at $(a/\sqrt{3},0,z_{A})$. It is easy to see that $U_{A}(a/\sqrt{3},0,z_{A})=-9V_{A}$ and $U_{B}(a/\sqrt{3},0,z_{A})=U_{C}(a/\sqrt{3},0,z_{A})=0$. Therefore, in contrast to $U_{A}$ which attains its minimum, $U_{B}$ and $U_{C}$ both attain their maxima at $(a/\sqrt{3},0,z_{A})$. To get an approximate model close to $(a/\sqrt{3},0,z_{A})$, we expand the optical dipole potentials of all three layers into the polynomials of the relative coordinates $\tilde{x}=x-a/\sqrt{3}$, $\tilde{y}=y$, and $\tilde{z}=z-z_{A}$. To the leading order of these relative coordinates, we have
\begin{equation}
\begin{cases}
U_A(\tilde{x},\tilde{y},\tilde{z})\simeq-9V_{A}+\frac{8\pi^{2}V_{A}}{a^{2}}(\tilde{x}^{2}+\tilde{y}^{2}) +\frac{18V_{A}}{\rho_{0z}^{2}}\tilde{z}^{2},  \\
U_B(\tilde{x},\tilde{y},\tilde{z})\simeq -\frac{4\pi^{2}V_{B}}{a^{2}}e^{-2z_{A}^{2}/\rho_{0z}^{2}}(\tilde{x}^{2}+\tilde{y}^{2}), \\
U_C(\tilde{x},\tilde{y},\tilde{z})\simeq -\frac{4\pi^{2}V_{C}}{a^{2}}e^{-2(z_{A}-z_{C})^{2}/\rho_{0z}^{2}}(\tilde{x}^{2}+\tilde{y}^{2}). \\
\end{cases}
\end{equation}
Therefore, for a potential well of $U_{A}$, the presence of $U_{B}$ and $U_{C}$ reduces the strength of the harmonic oscillator in the $xy$ plane from $\frac{8\pi^{2}V_{A}}{a^{2}}$ to
\begin{equation}
\frac{8\pi^{2}V_{A}}{a^{2}}[1-\frac{V_{B}}{2V_{A}}e^{-2z_{A}^{2}/\rho_{0z}^{2}} -\frac{V_{C}}{2V_{A}}e^{-2(z_{A}-z_{C})^{2}/\rho_{0z}^{2}}]\equiv\frac{8\pi^{2}\bar{V}_{A}}{a^{2}},
\end{equation}
where we have defined the reduced parameter $\bar{V}_{A}$. To the leading order of $\tilde{z}$, the harmonic oscillator along the $z$ direction is unchanged. Effectively, the presence of $U_B$ and $U_C$ enhances the relative strength of the approximate harmonic confinement along the $z$ direction. The ground state energy of this anisotropic harmonic oscillator model is
\begin{equation}
\varepsilon_{0A}=-9V_{A}+\hbar\omega_{xyA}+\frac{1}{2}\hbar\omega_{zA},
\end{equation}
where the angular frequencies are
\begin{equation}
\omega_{xyA}=\frac{4\pi}{a}\sqrt{\frac{\bar{V}_{A}}{m}}, \hspace{1cm} \omega_{zA}=\frac{6}{\rho_{0z}}\sqrt{\frac{V_{A}}{m}}.
\end{equation}

The analysis for the $B$ layer and the $C$ layer are the same. For the potential well of the B layer centering at $(0,0,0)$, the effective anisotropic harmonic oscillator potential is
\begin{equation}
U_B(\tilde{x},\tilde{y},\tilde{z})\simeq-9V_{B}+\frac{8\pi^{2}\bar{V}_{B}}{a^{2}}(\tilde{x}^{2}+\tilde{y}^{2}) +\frac{18V_{B}}{\rho_{0z}^{2}}\tilde{z}^{2},
\end{equation}
where the relative coordinates are simply $\tilde{x}=x$, $\tilde{y}=y$, and $\tilde{z}=z$. The reduced oscillator strength in the $xy$ plane is characterized by
\begin{equation}
\bar{V}_{B}=V_{B}[1-\frac{V_{A}}{2V_{B}}e^{-2z_{A}^{2}/\rho_{0z}^{2}}-\frac{V_{C}}{2V_{B}}e^{-2z_{C}^{2}/\rho_{0z}^{2}}].
\end{equation}
The ground state energy of this anisotropic harmonic oscillator model is
\begin{equation}
\varepsilon_{0B}=-9V_{B}+\hbar\omega_{xyB}+\frac{1}{2}\hbar\omega_{zB},
\end{equation}
where the angular frequencies are
\begin{equation}
\omega_{xyB}=\frac{4\pi}{a}\sqrt{\frac{\bar{V}_{B}}{m}}, \hspace{1cm} \omega_{zB}=\frac{6}{\rho_{0z}}\sqrt{\frac{V_{B}}{m}}.
\end{equation}
For the potential well of the C layer centering at $(-a/\sqrt{3},0,z_{C})$, the effective anisotropic harmonic oscillator potential is
\begin{equation}
U_C(\tilde{x},\tilde{y},\tilde{z})\simeq-9V_{C}+\frac{8\pi^{2}\bar{V}_{C}}{a^{2}}(\tilde{x}^{2}+\tilde{y}^{2}) +\frac{18V_{C}}{\rho_{0z}^{2}}\tilde{z}^{2},
\end{equation}
where the relative coordinates are $\tilde{x}=x+a/\sqrt{3}$, $\tilde{y}=y$, and $\tilde{z}=z-z_{C}$. The reduced oscillator strength in the $xy$ plane is characterized by
\begin{equation}
\bar{V}_{C}=V_{C}[1-\frac{V_{A}}{2V_{C}}e^{-2(z_{C}-z_{A})^{2}/\rho_{0z}^{2}}-\frac{V_{B}}{2V_{C}}e^{-2z_{C}^{2}/\rho_{0z}^{2}}].
\end{equation}
The ground state energy of this anisotropic harmonic oscillator model is
\begin{equation}
\varepsilon_{0C}=-9V_{C}+\hbar\omega_{xyC}+\frac{1}{2}\hbar\omega_{zC},
\end{equation}
where the angular frequencies are
\begin{equation}
\omega_{xyC}=\frac{4\pi}{a}\sqrt{\frac{\bar{V}_{C}}{m}}, \hspace{1cm} \omega_{zC}=\frac{6}{\rho_{0z}}\sqrt{\frac{V_{C}}{m}}.
\end{equation}


The wave function of the lowest bound state is written generally as
\begin{equation}
\psi_{0}(\mathbf{r}-\mathbf{r}_{0})=N_{0}e^{-\frac{\alpha_{xy}^{2}}{2}[(x-x_{0})^{2}+(y-y_{0})^{2}] -\frac{\alpha_{z}^{2}}{2}(z-z_{0})^{2}},
\end{equation}
where $\alpha_{xy}=\sqrt{m\omega_{xy}/\hbar}$, $\alpha_{z}=\sqrt{m\omega_{z}/\hbar}$, and the normalization factor $N_{0}=\alpha_{xy}\sqrt{\alpha_{z}}/\pi^{\frac{3}{4}}$. $m$, $a=\frac{2\lambda}{3}$, and $\rho_{0z}$ are the same for all three triangular layers. The angular frequencies are however dependent on the layer index through the above definitions. Correspondingly, according to the position of $\mathbf{r}_{0}$ and $\mathbf{r}_{1}$, we should put the layer index ($A$, $B$, or $C$) to $\omega_{xy}$, $\omega_{z}$, and other related quantities.


We are now ready to evaluate the hopping amplitudes.
For the single-body Hamiltonian defined by Eq.(E2) and the local orbital defined by Eq.(E18), the NN hopping amplitudes defined by Eq.(E5) may be calculated analytically. For the displaced dice lattice, there are six independent NN hopping amplitudes. For the intralayer NN hopping amplitude in the B layer, we take $\mathbf{r}_{0}=(0,0,0)$ and $\mathbf{r}_{1}=(\frac{\sqrt{3}}{2},\frac{1}{2},0)a$. The result for the integral is
\begin{eqnarray}
&&t_{NN}(BB)= \notag \\
&&-\frac{\hbar^{2}}{2m}(\frac{1}{4}\alpha_{xyB}^{4}a^{2}-\alpha_{xyB}^{2}-\frac{1}{2}\alpha_{zB}^{2})\exp(-\frac{1}{4}\alpha_{xyB}^{2}a^{2}) \notag \\
&&-3\{[1-\frac{2}{3}\exp(-\frac{b^{2}}{4\alpha_{xyB}^{2}})]V_{B}+[1+\frac{1}{3}\exp(-\frac{b^{2}}{4\alpha_{xyB}^{2}})]\cdot
\notag \\
&&\cdot[V_{A}\exp(-\frac{2\alpha_{zB}^{2}\rho_{0z}^{2}}{2+\alpha_{zB}^{2}\rho_{0z}^{2}}\frac{z_{A}^{2}}{\rho_{0z}^{2}})
\notag \\
&&+V_{C}\exp(-\frac{2\alpha_{zB}^{2}\rho_{0z}^{2}}{2+\alpha_{zB}^{2}\rho_{0z}^{2}}\frac{z_{C}^{2}}{\rho_{0z}^{2}})]\}\cdot
\notag \\
&&\cdot\sqrt{\frac{\alpha_{zB}^{2}\rho_{0z}^{2}}{2+\alpha_{zB}^{2}\rho_{0z}^{2}}}\exp(-\frac{1}{4}\alpha_{xyB}^{2}a^{2}).
\end{eqnarray}
A crucial qualitative feature of the result is the existence of a common exponential factor with the exponent proportional to the square of $a$, which is the distance between $\mathbf{r}_{0}$ and $\mathbf{r}_{1}$. Because the three layers of the displaced dice lattice are all triangular lattices and their relative positions are similar, we could get the NN hopping amplitudes within the A layer and those within the C layer by suitable substitutions of indices. Without writing out the explicit expressions, the relevant substitutions are simply
\begin{equation}
t_{NN}(AA)=t_{NN}(BB)|_{B\rightarrow A, A\rightarrow C, C\rightarrow B, z_{A}\rightarrow z_{C}-z_{A},z_{C}\rightarrow -z_{A}},
\end{equation}
\begin{equation}
t_{NN}(CC)=t_{NN}(BB)|_{B\rightarrow C, C\rightarrow A, A\rightarrow B, z_{A}\rightarrow -z_{C},z_{C}\rightarrow z_{A}-z_{C}}.
\end{equation}
Therefore, the NN intralayer hopping amplitudes in all three layers have an exponential factor, whose negative exponent is proportional to $a^{2}$, the square of the distance between the NN potential wells.

Now we calculate the three NN interlayer hopping amplitudes. For the NN hopping amplitude between the B layer and the A layer, we choose $\mathbf{r}_{0}=(0,0,0)$ on the B layer and $\mathbf{r}_{1}=(\frac{a}{\sqrt{3}},0,z_{A})$ on the A layer. The complete expression for the hopping amplitude is too lengthy and so we split it into four parts
\begin{equation}
t_{NN}(BA)=t^{K}_{BA}+t^{A}_{BA}+t^{B}_{BA}+t^{C}_{BA},
\end{equation}
where the first term is the matrix element of the kinetic energy term of $\hat{H}_{0}$, the remaining three terms are related separately to the three components of the potential energy term of $\hat{H}_{0}$. In addition, we define the following abbreviations for the composite quantities
\begin{equation}
\begin{cases}
\eta_{21}=\frac{2\alpha_{xyB}\alpha_{xyA}}{\alpha_{xyB}^{2}+\alpha_{xyA}^{2}},  \\
\zeta_{21}=\frac{2\alpha_{zB}\alpha_{zA}}{\alpha_{zB}^{2}+\alpha_{zA}^{2}},     \\
\alpha_{xy21}=\alpha_{xyB}\alpha_{xyA},     \\
\alpha_{z21}=\alpha_{zB}\alpha_{zA}.     \\
\end{cases}
\end{equation}
When $V_{A}$ and $V_{B}$ are at most slightly different (i.e., $V_{A}\simeq V_{B}$), which is the case that we focus on, $\eta_{21}\simeq 1$, $\zeta_{21}\simeq 1$, $\alpha_{xy21}\simeq \alpha_{xyB}^{2}$, and $\alpha_{z21}\simeq \alpha_{zB}^{2}$.
The results for the four terms of $t_{NN}(BA)$ are
\begin{eqnarray}
t^{K}_{BA}&=&-\frac{\hbar^{2}}{2m}\eta_{21}\sqrt{\zeta_{21}} \{\frac{(\eta_{21}\alpha_{xy21})^{2}}{4}(\frac{a}{\sqrt{3}})^{2}
\notag  \\
&&+\frac{(\zeta_{21}\alpha_{z21})^{2}}{4}z_{A}^{2}-\eta_{21}\alpha_{xy21}-\frac{\zeta_{21}\alpha_{z21}}{2}\}\cdot
\notag  \\
&&\cdot\exp[-\frac{\eta_{21}\alpha_{xy21}}{4}(\frac{a}{\sqrt{3}})^{2}-\frac{\zeta_{21}\alpha_{z21}}{4}z_{A}^{2}].
\end{eqnarray}
\begin{eqnarray}
t^{A}_{BA}&=&-3V_{A}\eta_{21}\sqrt{\frac{2\alpha_{z21}\rho_{0z}^{2}}{4+(\alpha_{zB}^{2}+\alpha_{zA}^{2})\rho_{0z}^{2}}}\cdot
\notag  \\
&&\cdot\{1+\frac{2}{3}e^{\frac{-b^{2}}{2(\alpha_{xyB}^{2}+\alpha_{xyA}^{2})}} [\cos\frac{4\pi\alpha_{xyB}^{2}}{3(\alpha_{xyB}^{2}+\alpha_{xyA}^{2})}
\notag  \\
&&+2\cos\frac{2\pi\alpha_{xyB}^{2}}{3(\alpha_{xyB}^{2}+\alpha_{xyA}^{2})}]\}\cdot
 \\
&&\cdot\exp[-\frac{\eta_{21}\alpha_{xy21}}{4}(\frac{a}{\sqrt{3}})^{2} -\frac{(4+\alpha_{zA}^{2}\rho_{0z}^{2})\alpha_{zB}^{2}}{8+2(\alpha_{zB}^{2}+\alpha_{zA}^{2})\rho_{0z}^{2}}z_{A}^{2}]. \notag
\end{eqnarray}
\begin{equation}
t^{B}_{BA}=t^{A}_{BA}|_{V_{A}\rightarrow V_{B},\alpha_{xyA}\leftrightarrow\alpha_{xyB},\alpha_{zA}\leftrightarrow\alpha_{zB}}.
\end{equation}
\begin{eqnarray}
t^{C}_{BA}&=&-3V_{C}\eta_{21}\sqrt{\frac{2\alpha_{z21}\rho_{0z}^{2}}{4+(\alpha_{zB}^{2}+\alpha_{zA}^{2})\rho_{0z}^{2}}}\cdot
\notag  \\
&&\cdot\{1+\frac{2}{3}e^{\frac{-b^{2}}{2(\alpha_{xyB}^{2}+\alpha_{xyA}^{2})}} [\cos\frac{2\pi(\alpha_{xyB}^{2}-\alpha_{xyA}^{2})}{3(\alpha_{xyB}^{2}+\alpha_{xyA}^{2})}
\notag  \\
&&+2\cos\frac{2\pi(\alpha_{xyB}^{2}+2\alpha_{xyA}^{2})}{3(\alpha_{xyB}^{2}+\alpha_{xyA}^{2})}]\}\cdot
\notag  \\
&&\cdot\exp[-\frac{\eta_{21}\alpha_{xy21}}{4}(\frac{a}{\sqrt{3}})^{2}
\\
&&-\frac{(\alpha_{z21}\rho_{0z}z_{A})^{2}+4(\alpha_{zB}z_{C})^{2}+4\alpha_{zA}^{2}(z_{C}-z_{A})^{2}}{8+2(\alpha_{zB}^{2}+\alpha_{zA}^{2})\rho_{0z}^{2}}]. \notag
\end{eqnarray}
In comparison to the three NN intralayer hopping amplitudes, the exponential factors have two changes. Firstly, the in-plane exponent is now proportional to $(\frac{a}{\sqrt{3}})^{2}$ instead of $a^{2}$. Secondly, there is additional exponential decay associated with the difference in the $z$-coordinates (i.e., depending on $z_{A}$ and $z_{C}$).

For the NN hopping amplitude between the B layer and the C layer, we choose $\mathbf{r}_{0}=(0,0,0)$ on the B layer and $\mathbf{r}_{1}=(-\frac{a}{\sqrt{3}},0,z_{C})$ on the C layer. The complete expression for the hopping amplitude is again split into four parts
\begin{equation}
t_{NN}(BC)=t^{K}_{BC}+t^{A}_{BC}+t^{B}_{BC}+t^{C}_{BC},
\end{equation}
where the first term is the matrix element of the kinetic energy term of $\hat{H}_{0}$, the remaining three terms are related separately to the three components of the potential energy term of $\hat{H}_{0}$. The four terms of $t_{NN}(BC)$ turn out to be very similar in expression to the four terms of $t_{NN}(BA)$, and could be obtained from the expressions shown above by the following substitutions
\begin{equation}
\begin{cases}
t^{K}_{BC}=t^{K}_{BA}|_{A\rightarrow C},   \\
t^{A}_{BC}=t^{C}_{BA}|_{A\rightarrow C,C\rightarrow A},   \\
t^{B}_{BC}=t^{B}_{BA}|_{A\rightarrow C},   \\
t^{C}_{BC}=t^{A}_{BA}|_{A\rightarrow C}.   \\
\end{cases}
\end{equation}
From this similarity, $t_{NN}(BC)$ show similar dependencies as $t_{NN}(BA)$ on the parameters of the optical dipole potential.

Finally, for the NN hopping amplitude between the A layer and the C layer, we choose $\mathbf{r}_{0}=(\frac{a}{\sqrt{3}},0,z_{A})$ on the A layer and $\mathbf{r}_{1}=(\frac{a}{2\sqrt{3}},\frac{a}{2},z_{C})$ on the C layer. The complete expression for the hopping amplitude is also split into four parts
\begin{equation}
t_{NN}(AC)=t^{K}_{AC}+t^{A}_{AC}+t^{B}_{AC}+t^{C}_{AC},
\end{equation}
where the first term is the matrix element of the kinetic energy term of $\hat{H}_{0}$, the remaining three terms are related separately to the three components of the potential energy term of $\hat{H}_{0}$. These four terms may also be obtained from the four terms for $t_{NN}(BA)$ by substitution of parameters as follows
\begin{equation}
\begin{cases}
t^{K}_{AC}=t^{K}_{BA}|_{B\rightarrow C,z_{A}\rightarrow z_{A}-z_{C}},   \\
t^{A}_{AC}=t^{A}_{BA}|_{B\rightarrow C,z_{A}\rightarrow z_{A}-z_{C}},   \\
t^{C}_{AC}=t^{A}_{AC}|_{A\leftrightarrow C}=t^{A}_{BA}|_{B\rightarrow A, A\rightarrow C,z_{A}\rightarrow z_{A}-z_{C}},  \\
t^{B}_{AC}=t^{C}_{BA}|_{V_{C}\rightarrow V_{B},\alpha_{xyB}\rightarrow\alpha_{xyC},\alpha_{zB}\rightarrow\alpha_{zC},z_{A}\leftrightarrow z_{A}-z_{C}}.   \\
\end{cases}
\end{equation}

In comparison with $t_{NN}(BA)$ and $t_{NN}(BC)$ whose dominant terms have an exponent proportional to $-z_{A}^{2}$ and $-z_{C}^{2}$, $t_{NN}(AC)$ is exponentially smaller because the dominant terms have an exponent proportional to $-(z_{A}-z_{C})^{2}$. For example, $(z_{A}-z_{C})^{2}=4z_{A}^{2}$ in the special case $z_{A}=-z_{C}$, which amount to a fourth-power decay in the hopping amplitude. On the other hand, we have shown that the interlayer NN hopping amplitudes have an exponent proportional to $(\frac{a}{\sqrt{3}})^{2}$ which is $\frac{1}{3}$ of the corresponding exponent for the three intralayer NN hopping amplitudes. This shows that the intralayer NN hopping amplitudes have a component that is third power smaller than the corresponding factor for the interlayer NN hopping amplitudes. Combining these two trends, it is possible to tune $z_{A}$ and $-z_{C}$ to certain medium values, so that all the four kinds of hopping amplitudes that are neglected in the tight-binding model are much smaller (e.g., smaller by about one to two orders of magnitude) than $t_{NN}(BA)$ and $t_{NN}(BC)$ that are retained in the model.

As an order-of-magnitude comparison among the six NN hopping amplitudes, let us set $V_{A}=V_{B}=V_{C}$ and $z_{A}=-z_{C}=z_{0}$. For simplicity, we also suppress the layer indices on $\alpha_{xy}$ and $\alpha_{z}$. The magnitudes of the various hopping amplitudes are determined by the dominant exponential factors therein. From the above explicit expressions, the dominant exponential factors are $\exp[-(\alpha_{xy}a)^{2}/4]$ for $t_{NN}(\alpha\alpha)$ ($\alpha=A,B,C$), are $\exp[-(\alpha_{xy}a)^{2}/12-(\alpha_{z}z_{0})^{2}/4]$ for $t_{NN}(BA)$ and $t_{NN}(BC)$, and is $\exp[-(\alpha_{xy}a)^{2}/12-(\alpha_{z}z_{0})^{2}]$ for $t_{NN}(AC)$. To justify the tight-binding model of Eq.(E1), we require the exponents for $t_{NN}(BA)$ and $t_{NN}(BC)$ to be at least two orders of magnitude larger than those for the other four NN hopping amplitudes. This amounts to
\begin{equation}
\exp[-\frac{(\alpha_{xy}a)^{2}}{6}+\frac{(\alpha_{z}z_{0})^{2}}{4}]<10^{-2},
\end{equation}
and
\begin{equation}
\exp[-\frac{3(\alpha_{z}z_{0})^{2}}{4}]<10^{-2}.
\end{equation}
Eq.(E33) leads to $\exp[-\frac{(\alpha_{z}z_{0})^{2}}{4}]<(0.01)^{1/3}\simeq0.2154$, and correspondingly $\exp[\frac{(\alpha_{z}z_{0})^{2}}{4}]>(100)^{1/3}\simeq4.642$. Substituting $\exp[\frac{(\alpha_{z}z_{0})^{2}}{4}]=5$ into Eq.(E32), we get $\exp[-\frac{(\alpha_{xy}a)^{2}}{12}]<\sqrt{0.002}\simeq0.0447$. This inequality sets a lower bound to the strength of the optical dipole potential, which turns out to be $\bar{V}_{\alpha}>1.004E_{R}$ or equivalently $\hbar\omega_{xy\alpha}>4.251E_{R}$ ($\alpha=A,B,C$). This constraint is weaker than that inferred in Sec.II for the 2D limit. It should always be satisfied for the considered deep lattices. For known strengths of the optical dipole potential and the $\rho_{0z}$ parameter, $\exp[\frac{(\alpha_{z}z_{0})^{2}}{4}]\simeq5$ or larger may easily be fulfilled by tuning $z_{0}$.

Suppose we have tuned the parameters so that the BA and BC NN interlayer hopping amplitudes dominate among the six NN hopping amplitudes. The tight-binding model has four parameters [$\varepsilon_{ab}$, $\varepsilon_{cb}$, $t_{ba}=t_{NN}(BA)$, and $t_{bc}=t_{NN}(BC)$]. Correspondingly, we also have control over four parameters of the optical dipole potential ($V_{A}/V_{B}$, $V_{C}/V_{B}$, $z_{A}-z_{B}=z_{A}$, $z_{C}-z_{B}=z_{C}$). Therefore, we may tune the four parameters in the tight-binding model freely within the scope of applicability of the model.

We mention in passing that, all the tight-binding parameters obtained above may be reexpressed in terms of strengths of the optical potential (i.e., $V_{i}$, $i=A,B,C$), the geometric parameters (i.e., $\rho_{0z}$ and $z_{i}$, $i=A,B,C$) measured in units of the wavelength $\lambda$, and the recoil energy $E_{R}$. In particular, it is conventional to express all the energies in units of the recoil energy. The conversion is easy to make through the following relations
\begin{equation}
\begin{cases}
\hbar\omega_{xyi}=3\sqrt{2}E_{R}\sqrt{\frac{\bar{V}_{i}}{E_{R}}},  \\
\hbar\omega_{zi}=\frac{3\sqrt{2}}{\pi\gamma_{0}}E_{R}\sqrt{\frac{V_{i}}{E_{R}}}, \\
\alpha_{xyi}^{2}\lambda^{2}=6\sqrt{2}\pi^{2}\sqrt{\frac{\bar{V}_{i}}{E_{R}}}, \\
\alpha_{zi}^{2}\lambda^{2}=\frac{6\sqrt{2}\pi}{\gamma_{0}}\sqrt{\frac{V_{i}}{E_{R}}}, \\
\end{cases}
\end{equation}
where $\gamma_{0}=\rho_{0z}/\lambda$, $i=A,B,C$. We will not write out the explicit expressions of the tight-binding parameters in terms of this alternative parameter set.

\section{Implementation of the lattice phase-modulation spectroscopy for the interband transitions}

The conventional stimulus for the interband transitions of electronic systems in a crystal is the optical field, the electric field component of which causes parity-changing interband excitations. Because the wave vector of the optical field triggering the excitation is much smaller than the scale of the wave vectors in the Brillouin zone (BZ) of the material, the optical excitations are usually taken as vertical. That is, the initial and final electronic states of the transition have the same wave vector. For a light field with the electric field component $\mathbf{E}(\mathbf{r},t)$, its interaction with the electronic system is written as
\begin{equation}
-q\mathbf{E}(\mathbf{r},t)\cdot\mathbf{r}=-\mathbf{F}_{EM}(\mathbf{r},t)\cdot\mathbf{r},
\end{equation}
where $q$ is the charge of the electron and $\mathbf{F}_{EM}(\mathbf{r},t)$ is the force acting on the electrons at the position $\mathbf{r}$ and time $t$. Only the electric field component is retained because the effect of the magnetic field component is negligible. For optical excitations of a 2D or q-2D material, it is usually reasonable to neglect the spatial dependence of the electric field and the force in all directions, and thus $\mathbf{F}_{EM}(\mathbf{r},t)=\mathbf{F}_{EM}(t)$.

For the neutral ultracold atoms on an optical lattice, a similar interaction may be constructed by virtue of the analogy between the electromagnetic Lorentz force and the inertial force for the center-of-mass motion \cite{tokuno11,wu15,anderson19,asteria19,drese97,eckardt17}. By shaking the optical lattice periodically, the optical dipole potential of the optical lattice oscillate in space in a periodic manner. We assume the modulation $\mathbf{l}(t)$ of the optical lattice (the optical potential) in the laboratory frame is uniform and therefore independent of $\mathbf{r}$. Turning to the reference frame comoving with the optical lattice, all ultracold atoms in the optical lattice are subjected to the following inertial force
\begin{equation}
\mathbf{F}_{inertial}(\mathbf{r},t)=\mathbf{F}_{inertial}(t)=-m\frac{d^{2}\mathbf{l}(t)}{dt^{2}}\equiv-m\ddot{\mathbf{l}}.
\end{equation}
This inertial force in turn amounts to adding the following potential energy to the model for the cold atoms in the comoving frame
\begin{equation}
\hat{H}_{d}=-\mathbf{F}_{inertial}(t)\cdot\mathbf{r}.
\end{equation}
For a sinusoidally shaken optical lattice, the above formula has the same form as Eq.(F1) for the electronic systems. When transformed to the tight-binding model, we replace $\mathbf{r}$ by the position operator of the cold atoms. Therefore, sinusoidally shaking the optical lattice may excite the parity-changing interband excitations in the cold atoms, just as the optical field excites the same kind of excitations in electronic systems \cite{tokuno11,wu15,anderson19,asteria19,drese97,eckardt17}.

In addition to the above heuristic justifications, the interaction term may be derived from a more formal approach \cite{tokuno11,drese97,eckardt17}. For this purpose, we specify the explicit expression of the sinusoidal modulation $\mathbf{l}(t)$. For generality, we consider an elliptical shaking in the $xy$ plane of the optical lattice. That is, in each period of revolution, $\mathbf{l}(t)$ traces out an ellipse centering at $\mathbf{l}(t)=\mathbf{0}$. We assume the major (minor) axis of the ellipse is along $\mathbf{e}_{a}$ ($\mathbf{e}_{b}$), with $\mathbf{e}_{a}\cdot\mathbf{e}_{a}=\mathbf{e}_{b}\cdot\mathbf{e}_{b}=1$, $\mathbf{e}_{a}\cdot\mathbf{e}_{b}=0$, and $\mathbf{e}_{a}\times\mathbf{e}_{b}=\hat{\mathbf{z}}$. We have
\begin{equation}
\mathbf{l}(t)=a\cos(\omega_{0}t)\mathbf{e}_{a}+b\eta\sin(\omega_{0}t)\mathbf{e}_{b},
\end{equation}
where $\omega_{0}$ is the angular frequency of the shaking. $a\geqslant b\geqslant0$ characterize the amplitude of the driving. $\eta=\pm1$ characterizes the rotating direction (i.e., chirality) of $\mathbf{l}(t)$. Viewing against the $\hat{\mathbf{z}}$ direction, the rotation is counterclockwise (clockwise) for $\eta=1$ ($\eta=-1$). When $b=0$, the modulation is linear. When $a=b$, the modulation is circular. In terms of the following basis set
\begin{equation}
\begin{cases}
\mathbf{e}_{1}=\frac{1}{\sqrt{2}}(\mathbf{e}_{a}-\mathbf{e}_{b}),   \\
\mathbf{e}_{2}=\frac{1}{\sqrt{2}}(\mathbf{e}_{a}+\mathbf{e}_{b}),   \\
\end{cases}
\end{equation}
the sinusoidal modulation is reformulated as \cite{jotzu14}
\begin{equation}
\mathbf{l}(t)=c[\cos(\omega_{0}t+\varphi)\mathbf{e}_{1}+\cos(\omega_{0}t-\varphi)\mathbf{e}_{2}],
\end{equation}
where
\begin{equation}
\begin{cases}
c=\sqrt{(a^{2}+b^{2})/2},   \\
\sin\varphi=b\eta/(\sqrt{2}c),   \\
\cos\varphi=a/(\sqrt{2}c).   \\
\end{cases}
\end{equation}
Finally, we may of course express $\mathbf{l}(t)$ in terms of the basis vectors of the $Oxyz$ Cartesian coordinate system defined referring to the underlying optical lattice. Suppose the major axis (minor axis) of the ellipse is rotated from the $x$ axis ($y$ axis) by an angle of $\alpha$, we have
\begin{equation}
\begin{cases}
\mathbf{e}_{a}=\hat{\mathbf{x}}\cos\alpha+\hat{\mathbf{y}}\sin\alpha,   \\
\mathbf{e}_{b}=-\hat{\mathbf{x}}\sin\alpha+\hat{\mathbf{y}}\cos\alpha.   \\
\end{cases}
\end{equation}
In the coordinate basis, we express the modulation as
\begin{equation}
\mathbf{l}(t)=\delta x(t)\hat{\mathbf{x}}+\delta y(t)\hat{\mathbf{y}}.
\end{equation}
In the special cases with $\sin\alpha=0$ or $\cos\alpha=0$, $\delta x(t)$ and $\delta y(t)$ are very simply related to $l_{a}(t)$ and $l_{b}(t)$ defined in Eq.(F4). In other more general cases, we have
\begin{equation}
\begin{cases}
\delta x(t)=c_{1}\cos(\omega_{0}t+\varphi_{1}),  \\
\delta y(t)=c_{2}\cos(\omega_{0}t+\varphi_{2}),  \\
\end{cases}
\end{equation}
where we have defined
\begin{equation}
\begin{cases}
c_{1}=\sqrt{a^{2}\cos^{2}\alpha+b^{2}\sin^{2}\alpha},   \\
\cos\varphi_{1}=a\cos\alpha/c_{1},   \\
\sin\varphi_{1}=b\eta\sin\alpha/c_{1},   \\
\end{cases}
\end{equation}
and
\begin{equation}
\begin{cases}
c_{2}=\sqrt{a^{2}\sin^{2}\alpha+b^{2}\cos^{2}\alpha},   \\
\cos\varphi_{2}=a\sin\alpha/c_{2},   \\
\sin\varphi_{2}=-b\eta\cos\alpha/c_{2}.   \\
\end{cases}
\end{equation}
This general expression in the basis of $\hat{\mathbf{x}}$ and $\hat{\mathbf{y}}$ is clearly more complex and less comprehensible than either Eq.(F4) or Eq.(F6). Physically, however, all three expressions for $\mathbf{l}(t)$ are completely equivalent.

In the presence of the above elliptical modulation $\mathbf{l}(t)$, the optical dipole potential $V(\mathbf{r})$ becomes $\tilde{V}(\mathbf{r},t)=V(\mathbf{r}-\mathbf{l}(t))$. The single-body Hamiltonian for the ultracold atoms in the shaken optical lattice turns into
\begin{equation}
\hat{\tilde{H}}=\frac{\mathbf{p}^{2}}{2m}+\tilde{V}(\mathbf{r},t).
\end{equation}
In terms of two successive unitary transformations to the Schr\"{o}dinger equation \cite{drese97}, the time dependence of the model may be relegated from the potential energy term to a new term of the same form as Eq.(F3). We start from the Schr\"{o}dinger equation,
\begin{equation}
i\hbar\frac{\partial\tilde{\psi}(\mathbf{r},t)}{\partial t}=\hat{\tilde{H}}\tilde{\psi}(\mathbf{r},t).
\end{equation}
In the first transformation, we take
\begin{equation}
\tilde{\psi}(\mathbf{r},t)= e^{\frac{i}{\hbar}(-\mathbf{l}\cdot\mathbf{p}+\frac{m}{4}\mathbf{l}\cdot\dot{\mathbf{l}})}\tilde{\psi}_{1}(\mathbf{r},t) \equiv U_{1}(\mathbf{r},t)\tilde{\psi}_{1}(\mathbf{r},t).
\end{equation}
The factor $\exp(-i\mathbf{l}\cdot\mathbf{p}/\hbar)$ of $U_{1}$ implements a spatial translation of distance $\mathbf{l}$. The Hamiltonian becomes
\begin{eqnarray}
&&\hat{\tilde{H}}_{1}=U_{1}^{\dagger}\hat{\tilde{H}}U_{1}-i\hbar U_{1}^{\dagger}\frac{\partial}{\partial t}U_{1}  \notag \\
&&\hspace{0.5cm} =\frac{(\mathbf{p}-m\dot{\mathbf{l}})^{2}}{2m}+V(\mathbf{r})-\frac{1}{2}m\omega_{0}^{2}c^{2},
\end{eqnarray}
where $c=\sqrt{(a^{2}+b^{2})/2}$ is defined in Eq.(F7). In the second transformation, we take
\begin{equation}
\tilde{\psi}_{1}(\mathbf{r},t)= e^{\frac{i}{\hbar}m\dot{\mathbf{l}}\cdot\mathbf{r}}\tilde{\psi}_{2}(\mathbf{r},t) \equiv U_{2}(\mathbf{r},t)\tilde{\psi}_{2}(\mathbf{r},t).
\end{equation}
The Hamiltonian transforms into
\begin{eqnarray}
&&\hat{\tilde{H}}_{2}=U_{2}^{\dagger}\hat{\tilde{H}}_{1}U_{2}-i\hbar U_{2}^{\dagger}\frac{\partial}{\partial t}U_{2}  \notag \\
&&\hspace{0.5cm} =\frac{\mathbf{p}^{2}}{2m}+V(\mathbf{r})+m\ddot{\mathbf{l}}\cdot\mathbf{r}-\frac{1}{2}m\omega_{0}^{2}c^{2}.
\end{eqnarray}
It is easy to see that, the third term in the final expression of $\hat{\tilde{H}}_{2}$ is exactly the same as Eq.(F3).

How could we realize the above sinusoidal modulation in our displaced dice lattice, in experiments? Firstly, we express the general elliptical modulation defined above in terms of the parameters defining our optical lattice. From Eq.(4) of the main text, the positions of the potential minima of the optical lattice are controlled by the phase parameters $\phi_{i}^{(j)}$ ($i,j=1,2,3$). Therefore, $\mathbf{l}(t)$ is expressed in terms of the modulations of these phases (i.e., $\delta\phi_{i}^{(j)}$). Since all three layers are modulated in the same manner, the modulations in these phases are independent of the layer index $i$. We therefore define ($i,j=1,2,3$)
\begin{equation}
\delta\phi_{i}^{(j)}(t)=\delta\phi^{(j)}(t).
\end{equation}
From Eq.(4) of the main text, the displacement of the optical lattice in the $xy$ plane is related to these phase modulations through
\begin{equation}
\begin{cases}
\delta x(t)=-\frac{\sqrt{3}a}{4\pi}[\delta\phi^{(1)}(t)-\delta\phi^{(3)}(t)],   \\
\delta y(t)=\frac{a}{4\pi}[\delta\phi^{(1)}(t)+\delta\phi^{(3)}(t)-2\delta\phi^{(2)}(t)],   \\
\end{cases}
\end{equation}
where $a=2\lambda/3$ is the lattice constant of the displaced dice lattice. We assume $\delta\phi^{(j)}(t)\ll2\pi$ ($j=1,2,3$). Therefore, $\delta x(t)$ and $\delta y(t)$ are much smaller than the lattice constant $a$, corresponding to a weak stimulus to the system. There are infinitely many approaches of realizing the two displacements $\delta x(t)$ and $\delta y(t)$ in terms of three phase modulations $\delta\phi^{(j)}(t)$ ($j=1,2,3$). One implementation is by setting
\begin{equation}
\begin{cases}
\delta\phi^{(1)}(t)=\frac{4\pi}{\sqrt{3}a}[-\frac{1}{2}\delta x(t)+\frac{\sqrt{3}}{2}\delta y(t)],   \\
\delta\phi^{(2)}(t)=0,   \\
\delta\phi^{(3)}(t)=\frac{4\pi}{\sqrt{3}a}[\frac{1}{2}\delta x(t)+\frac{\sqrt{3}}{2}\delta y(t)].   \\
\end{cases}
\end{equation}
The coefficient $\frac{4\pi}{\sqrt{3}a}$ is the magnitude of the reciprocal lattice vector. In terms of Eqs.(F9)-(F12) and defining
\begin{equation}
\begin{cases}
C_{\zeta}=\frac{1}{2}\sqrt{c_{1}^{2}+3c_{2}^{2}+2\sqrt{3}\zeta c_{1}c_{2}\cos(\varphi_{1}-\varphi_{2})},   \\
\cos\varphi_{\zeta}=\frac{1}{2C_{\zeta}}(\zeta c_{1}\cos\varphi_{1}+\sqrt{3}c_{2}\cos\varphi_{2}),   \\
\sin\varphi_{\zeta}=\frac{1}{2C_{\zeta}}(\zeta c_{1}\sin\varphi_{1}+\sqrt{3}c_{2}\sin\varphi_{2}),   \\
\end{cases}
\end{equation}
where $\zeta=\pm$, we can reformulate Eq.(F21) as
\begin{equation}
\begin{cases}
\delta\phi^{(1)}(t)=\frac{4\pi C_{-}}{\sqrt{3}a}\cos(\omega_{0}t+\varphi_{-}),   \\
\delta\phi^{(2)}(t)=0,   \\
\delta\phi^{(3)}(t)=\frac{4\pi C_{+}}{\sqrt{3}a}\cos(\omega_{0}t+\varphi_{+}).   \\
\end{cases}
\end{equation}
An efficient and successful method of implementing such phase modulations is to attach piezoelectric actuators to certain optical elements related to the phase factors \cite{jotzu14,graham92}. According to Fig.3(a) of the main text, the $\delta\phi^{(1)}(t)$ [$\delta\phi^{(3)}(t)$] modulation may be realized by mounting the two mirrors M$_{3}$ and M$_{4}$ (M$_{1}$ and M$_{2}$) on piezoelectric actuators, and setting the motion of the two in step with each other \cite{jotzu14,graham92}.

Finally, we list the physical quantities relevant to the interband transitions of fermionic cold atoms confined in the symmetrically biased displaced dice lattice. For the circular modulation that we considered in the main text, we have $a=b$ and
\begin{equation}
\mathbf{l}(t)=a[\cos(\omega_{0}t)\hat{\mathbf{x}}+\eta\sin(\omega_{0}t)\hat{\mathbf{y}}].
\end{equation}
Similar to the polarized light field, it is more convenient to express the harmonically varying field in the complex representation, which brings the above expression to
\begin{equation}
\tilde{\mathbf{l}}(t)=a(\hat{\mathbf{x}}+\eta e^{i\frac{\pi}{2}}\hat{\mathbf{y}})e^{-i\omega_{0}t}=a(\hat{\mathbf{x}}+i\eta \hat{\mathbf{y}})e^{-i\omega_{0}t}.
\end{equation}
In the basis of $\hat{\mathbf{x}}$ and $\hat{\mathbf{y}}$, this gives the standard polarization vector (i.e., the Jones vector) $\mathbf{n}_{\eta}=(1,i\eta)/\sqrt{2}$ for the circularly polarized modulation. This modulation exerts a parity-changing perturbation that may trigger an interband transition. The strength of this transition is characterized by the matrix element of the component of the current operator (for the center-of-mass motion of the ultracold atoms) along $\dot{\mathbf{l}}(t)$ [$\dot{\tilde{\mathbf{l}}}(t)$]. According to Eq.(F16), $\dot{\mathbf{l}}(t)$ [$\dot{\tilde{\mathbf{l}}}(t)$] plays the role of the artificial vector potential. By transforming the Hamiltonian to the wave vector space, the $\mathbf{k}$-resolved current operator is represented as
\begin{equation}
\mathbf{J}(\mathbf{k})=\boldsymbol{\nabla}_{\mathbf{k}}h(\mathbf{k}).
\end{equation}
Therefore, the intensity of the vertical transition at $\mathbf{k}$ to a circular modulation characterized by $\mathbf{n}_{\eta}=(1,i\eta)/\sqrt{2}$ is proportional to the matrix element of
\begin{equation}
J_{\eta}(\mathbf{k})=\mathbf{J}(\mathbf{k})\cdot\mathbf{n}_{\eta}=\frac{1}{\sqrt{2}}[\frac{\partial h(\mathbf{k})}{\partial k_{x}}+i\eta\frac{\partial h(\mathbf{k})}{\partial k_{y}}].
\end{equation}
In the basis $\psi^{\dagger}_{\mathbf{k}\sigma}=[a^{\dagger}_{\mathbf{k}\sigma}, b^{\dagger}_{\mathbf{k}\sigma}, c^{\dagger}_{\mathbf{k}\sigma}]$, the tight-binding model in the $\mathbf{k}$ space is
\begin{equation}
\hat{H}=\sum\limits_{\mathbf{k},\sigma}\psi^{\dagger}_{\mathbf{k}\sigma}h(\mathbf{k})\psi_{\mathbf{k}\sigma}.
\end{equation}
The Hamiltonian matrix that is independent of the spin label is
\begin{equation}
h(\mathbf{k})=\begin{pmatrix} \varepsilon_{ab} & \xi_{ab}(\mathbf{k}) & 0  \\
                              \xi_{ab}^{\ast}(\mathbf{k}) & 0 & \xi_{bc}(\mathbf{k}) \\
                              0 & \xi_{bc}^{\ast}(\mathbf{k}) & \varepsilon_{cb}
               \end{pmatrix},
\end{equation}
where
\begin{equation}
\begin{cases}
\xi_{ab}(\mathbf{k})=t_{ba}(1+e^{i\mathbf{k}\cdot\mathbf{a}_{1}}+e^{i\mathbf{k}\cdot\mathbf{a}_{2}}),  \\
\xi_{bc}(\mathbf{k})=t_{bc}(1+e^{i\mathbf{k}\cdot\mathbf{a}_{1}}+e^{i\mathbf{k}\cdot\mathbf{a}_{2}}).  \\
\end{cases}
\end{equation}
For $\varepsilon_{ab}=-\varepsilon_{cb}=\Delta$ and $t_{ba}=t_{bc}=t_{0}$, the above model reduces to the biased and displaced dice lattice that we studied in the main text. As was shown in the main text, the band structure of this model has three bands. Each band is twofold degenerate by spin. We suppress the spin label hereafter since it is irrelevant to the present discussions and will only give a factor of two to the final results.

We take the normalized eigenvector for the flat band at energy $E_{0}(\mathbf{k})=0$ as
\begin{equation}
\langle\psi_{0}(\mathbf{k})|=\frac{1}{E(\mathbf{k})}\begin{pmatrix} -\xi^{\ast}(\mathbf{k}), & \Delta, & \xi(\mathbf{k}) \end{pmatrix},
\end{equation}
where $\xi(\mathbf{k})=t_{0}(1+e^{i\mathbf{k}\cdot\mathbf{a}_{1}}+e^{i\mathbf{k}\cdot\mathbf{a}_{2}})$.
For the two dispersive bands with energies $E_{\nu}(\mathbf{k})=\nu E(\mathbf{k})$ ($\nu=\pm$), we take
\begin{equation}
\langle\psi_{\nu}(\mathbf{k})|=\frac{|\xi(\mathbf{k})|}{E(\mathbf{k})}
\begin{pmatrix} \frac{\xi^{\ast}(\mathbf{k})}{E(\mathbf{k})-\nu\Delta}, & \nu, & \frac{\xi(\mathbf{k})}{E(\mathbf{k})+\nu\Delta} \end{pmatrix}.
\end{equation}
The three components of the eigenvectors are the amplitudes of the eigenvectors on the three sublattices.

Substituting $h(\mathbf{k})$ and its eigenvectors defined above to Eq.(6) of the main text, we obtain the valley-contrasting interband excitations shown in Eq.(7) of the main text.

It is interesting to compare the above valley-contrasting interband transitions with the interband transitions for two closely related situations. In the first case, the transitions are still for the symmetrically biased dice model, but from the lower dispersive band to the upper dispersive band. In terms of Eq.(6) of the main text, now we have $\psi_{i}=\psi_{-}$ and $\psi_{f}=\psi_{+}$. The resonance frequency (i.e., $\hbar\omega_{0}'=2|\Delta|$) for this transition is twice the resonance frequency (i.e., $\hbar\omega_{0}=|\Delta|$) for the transition between the flat band and one dispersive band. Substituting the above definitions to Eq.(6) of the main text, the matrix element for this transition is
\begin{equation}
\langle\psi_{+}(\mathbf{k})|[\frac{\partial h(\mathbf{k})}{\partial k_{x}}+i\eta\frac{\partial h(\mathbf{k})}{\partial k_{y}}]
|\psi_{-}(\mathbf{k})\rangle=0.
\end{equation}
Therefore, the direct transitions between the two dispersive bands are forbidden under a sinusoidal shaking of the optical lattice.

In the second case, we consider the interband transitions for the model with parameters $\varepsilon_{ab}=\varepsilon_{cb}=\Delta$ and $t_{ba}=t_{bc}=t_{0}$. This variant of the dice model may also be realized by the more conventional method for the optical lattice \cite{rizzi06,bercioux09}. Because this model has the inversion symmetry, the interband transitions in this model are expected to be non-valley-contrasting \cite{yao08,xiao12,cao12}. This model also has one flat band and two dispersive bands. The flat band $E'_{0}(\mathbf{k})=\Delta$. The dispersions of the other two bands are
\begin{equation}
E'_{\nu}(\mathbf{k})=\frac{1}{2}[\Delta+\nu\sqrt{\Delta^{2}+8|\xi(\mathbf{k})|^{2}}],
\end{equation}
where the band index $\nu=\pm$. In this model, the flat band is not completely isolated. Instead, the flat band connects to the $\nu$th dispersive band satisfying $\nu\Delta/|\Delta|=1$. The eigenvector for the flat band is taken as
\begin{equation}
\langle\psi'_{0}(\mathbf{k})|=\frac{1}{\sqrt{2}}\begin{pmatrix} \frac{\xi^{\ast}(\mathbf{k})}{|\xi(\mathbf{k})|}, & 0, & -\frac{\xi(\mathbf{k})}{|\xi(\mathbf{k})|} \end{pmatrix},
\end{equation}
and for the two dispersive bands with energies $E'_{\nu}(\mathbf{k})$ ($\nu=\pm$) as
\begin{equation}
\langle\psi'_{\nu}(\mathbf{k})|=\sqrt{\frac{E'^{2}_{\nu}(\mathbf{k})}{E'^{2}_{\nu}(\mathbf{k})+2|\xi(\mathbf{k})|^{2}}}
\begin{pmatrix} \frac{\xi^{\ast}(\mathbf{k})}{E'_{\nu}(\mathbf{k})}, & 1, & \frac{\xi(\mathbf{k})}{E'_{\nu}(\mathbf{k})} \end{pmatrix}.
\end{equation}
Again, we consider the vertical interband transitions under a resonant circular driving. The amplitude of the transition at $\mathbf{k}$ is determined by the following matrix element
\begin{equation}
\langle\psi'_{f}(\mathbf{k})|J_{\eta}(\mathbf{k})
|\psi'_{i}(\mathbf{k})\rangle,
\end{equation}
where $\eta=\pm$, the current operator for the present model is the same as that for the symmetrically biased dice model. Introducing the relative momenta $\mathbf{q}=\mathbf{k}-\mathbf{K}_{\tau}$, where $\tau=\pm$, $\mathbf{K}_{+}=\mathbf{K}^{\prime}$, and $\mathbf{K}_{-}=\mathbf{K}$, and expanding the matrix element to the leading order of $\mathbf{q}$, we have
\begin{eqnarray}
&&\langle\psi'_{0}(\mathbf{K}_{\tau}+\mathbf{q})|J_{\eta}(\mathbf{K}_{\tau}+\mathbf{q})
|\psi'_{\nu}(\mathbf{K}_{\tau}+\mathbf{q})\rangle    \notag \\
&&\simeq -\frac{\sqrt{3}a}{4}|t_{0}|[1-\nu\text{sgn}(\Delta)]\tau\frac{q_{x}\eta+iq_{y}}{q},
\end{eqnarray}
where $q=\sqrt{q_{x}^{2}+q_{y}^{2}}$. There are two qualitative differences from the corresponding transitions in the symmetrically biased dice model. Firstly, the transition between the flat band and the band connecting to it vanishes close to the touching points $\mathbf{K}_{\tau}$ ($\tau=\pm$). The nonvanishing transition at the band edge is resonant when the modulation frequency satisfies $\hbar\omega_{0}=|\Delta|$. Secondly, the transition is not valley contrasting, in agreement with the presence of inversion symmetry.

Another difference compared to the symmetrically biased dice model is that the interband transitions between the two dispersive bands are nonvanishing. The corresponding matrix element is
\begin{eqnarray}
&&\langle\psi'_{+}(\mathbf{K}_{\tau}+\mathbf{q})|J_{\eta}(\mathbf{K}_{\tau}+\mathbf{q})
|\psi'_{-}(\mathbf{K}_{\tau}+\mathbf{q})\rangle      \notag \\
&&\simeq  -\frac{\sqrt{3}a}{2}|t_{0}|\text{sgn}(\Delta)\frac{q_{x}+i\eta q_{y}}{q}(1-i\frac{\sqrt{3}a}{2}q_{x}\eta\tau).
\end{eqnarray}
This transition is also not valley contrasting. In addition, same as the above nonvanishing transition involving the flat band, the present transition close to the band edge is also resonant at the modulation frequency $\hbar\omega_{0}=|\Delta|$. Therefore, two types of interband transitions occur simultaneously in this model.

\end{appendix}


\end{document}